\documentclass[twocolumn]{aastex701}

\usepackage{graphicx}
\usepackage{amsmath}
\usepackage{float}
\usepackage{capt-of}
\usepackage{placeins}
\usepackage{cancel}

\usepackage[utf8]{inputenc}

\begin{document}

\title{The IRAC Distant Cluster Survey}

\author[orcid=0000-0001-7145-549X,sname='Garcia']{Karolina Garcia}
\affiliation{Center for AstroPhysical Surveys, National Center for Supercomputing Applications, University of Illinois,\\1205 West Clark Street, Urbana, IL 61801, USA}
\affiliation{Department of Astronomy, University of Illinois, 1002 W Green St, Urbana, IL 61801, USA}
\affiliation{Department of Astronomy, University of Florida, 211 Bryant Space Science Center, Gainesville, FL 32611, USA}
\email[show]{karolinagarcia.contact@gmail.com}  

\author[0000-0002-0933-8601]{Anthony H. Gonzalez}
\affil{Department of Astronomy, University of Florida, 211 Bryant Space Science Center, Gainesville, FL 32611, USA}
\email{.}  

\author[0000-0002-4208-798X]{Mark Brodwin}
\affiliation{Eureka Scientific, 2452 Delmer Street Suite 100, Oakland, CA, 94602-3017, USA}
\email{.}  

\author{Peter Eisenhardt}
\affiliation{Jet Propulsion Laboratory, California Institute of Technology, 4800 Oak Grove Dr., Pasadena, CA 91109, USA}
\email{.}  

\author[0000-0001-6654-6762]{Tyler George}
\affil{Department of Physics, Stony Brook University, 100 Nicolls Road, Stony Brook, NY 11790, USA}
\email{.}  

\author[0009-0009-2957-3033]{Nadia Hansen}
\affil{Department of Astronomy, University of Florida, 211 Bryant Space Science Center, Gainesville, FL 32611, USA}
\email{.}  

\author[0000-0003-0122-0841]{S. A.\ Stanford}
\affiliation{Department of Physics \& Astronomy, University of California, One Shields Avenue, Davis, CA 95616, USA}
\email{.}

\author[0000-0003-2686-9241]{Daniel Stern}
\affiliation{Jet Propulsion Laboratory, California Institute of Technology, 4800 Oak Grove Dr., Pasadena, CA 91109, USA}
\email{.}  

\author[0000-0001-7027-2202]{Khunanon Thongkham}
\affil{Korea Astronomy and Space Science Institute, 776, Daedeokdae-ro, Yuseong-gu, Daejeon 34055, Republic of Korea}
\affil{National Astronomical Research Institute of Thailand (NARIT), Mae Rim, Chiang Mai 50180, Thailand}
\affil{Department of Astronomy, University of Florida, 211 Bryant Space Science Center, Gainesville, FL 32611, USA}
\email{.}  

\author[0000-0003-3428-1106]{Ariane Trudeau}
\affil{Institute of Astronomy and Astrophysics, Academia Sinica, Taipei 10617, Taiwan}
\affil{Department of Astronomy, University of Florida, 211 Bryant Space Science Center, Gainesville, FL 32611, USA}
\email{.}

\author[0000-0002-1207-9137]{Michael J. I. Brown}
\affil{School of Physics \& Astronomy, Monash University, Clayton, VIC 3800, Australia}
\email{.}

\author[0000-0002-1578-6582]{Buell T. Jannuzi}
\affil{Steward Observatory, University of Arizona, 933 North Cherry Avenue, Tucson, AZ 85721, USA}
\email{.}

\begin{abstract}
We present the Infrared Array Camera (IRAC) Distant Cluster Survey (IDCS), a catalog of 825 galaxy cluster and group candidates identified in the Boötes field using the \texttt{PZWav} detection code. Candidates are selected using photometric redshift probability distributions derived from twelve-band optical and infrared imaging. The catalog spans $0.08 < z < 2.0$ (median $z = 0.93$) and includes 147 systems at $z \geq 1.5$. For $z < 1.5$, simulations indicate a purity of $93\%$ at the adopted S/N $\geq 4$ threshold and $50\%$ completeness at $\log(M_{200c}/M_\odot) = 13.74 \pm 0.01$. We validate the cluster photometric redshifts through comparison with spectroscopic data, finding $\sigma_z/(1+z)=0.017$. Including 18 previously published IRAC Shallow Cluster Survey (ISCS) systems, a total of 80 IDCS cluster candidates are now confirmed through spectroscopy and/or X-ray observations. DESI DR1 spectroscopy provides redshifts for 38 systems, including 23 newly confirmed clusters. These DESI-confirmed systems span $0.12 < z < 0.93$ and have at least 10 spectroscopic members each, from which we measure a median velocity dispersion of $\sim 490~\mathrm{km\,s^{-1}}$, corresponding to a median halo mass of $1.1 \times 10^{14}\,M_\odot$. A positional comparison with 39 independently confirmed literature clusters yields a median centroid offset of 211~kpc. The IDCS catalog provides a cluster sample spanning a wide redshift range, with a selection function characterized through simulations and is a valuable resource for studies of large-scale structure, galaxy evolution in dense environments, and the calibration of cluster detection techniques for current and upcoming wide-field surveys, including LSST, Roman, and Euclid.
\end{abstract}


\section{Introduction} \label{sec:intro}

We are entering an era where wide-area surveys are capable of discovering large samples of galaxy clusters extending from the local universe out to the epoch of cluster assembly at $z\sim2$. Facilities including the Rubin Observatory \citep{ivezic_lsst_2019}, the Nancy Grace Roman Space Telescope \citep{spergel_wide-field_2015} and Euclid \citep{euclid_collaboration_euclid_2025} expect to discover hundreds of thousands to millions of clusters spanning this redshift range. Meanwhile, cosmic microwave background (CMB) surveys with facilities such as the Atacama Cosmology Telescope \citep[ACT,][]{act_2020,act_clusters_2021}, the South Pole Telescope \citep[SPT-3G,][]{SPT-3G,spt3g_clusters_2025,bleem_2026}, the Simons Observatory \citep{Simons_Obs}, the CMB-S4 experiment \citep{abazajian_cmb-s4_2019}, and deep XMM surveys such as FornaX \citep{fornax} have started surveying the intracluster medium (ICM) extending over this redshift range.

The Infrared Array Camera (IRAC) Shallow Cluster Survey \citep[ISCS,][]{iscs_paper} was among the first large tomographic surveys to identify galaxy clusters out to high redshift, using photometric redshifts to detect overdensities in redshift slices. Recent advances in photometric surveys and cluster identification techniques have greatly expanded the detection and characterization of galaxy clusters, particularly through large-area, multi-wavelength data sets. For example, the series of cluster catalogs by \citet{wen_2009, wen_2017} and \citet{wen_2011, wen_2021} identified a total of 110{,}621 clusters across several wide-area optical surveys. This growth has accelerated with surveys based on the Wide-field Infrared Survey Explorer (WISE) and the Dark Energy Camera (DECam). The second release of the Massive and Distant Clusters of WISE Survey \citep[MaDCoWS2;][]{thongkham_massive_2024_b,thongkham_massive_2024_a} reported $133{,}036$ candidates at $0.08<z<2$ over $6400$ deg$^2$, while \cite{wen_clusters_2022} found $151{,}244$ candidates up to $z=1.5$.

Most recently, \cite{WH2024} produced a catalog of 1.58 million clusters using stellar-mass overdensities in the Dark Energy Spectroscopic Instrument (DESI) Legacy Imaging Surveys \citep{dey_legacy_2019}. These unprecedentedly large samples offer the potential for refining cosmological constraints (e.g., \citeauthor{sartoris_next_2016}~\citeyear{sartoris_next_2016} and references therein) and for quantitatively assessing the impact of environmental quenching processes over cosmic time. However, a key challenge remains in characterizing the cluster catalogs generated by these extensive surveys. For these catalogs to support cosmological and environmental studies, both their selection functions and mass calibration must be well quantified, including completeness and contamination as functions of mass and redshift, the mass-observable relation, and the projection and photometric-redshift effects that bias these quantities.

A complementary approach is to leverage cluster catalogs from deep, well-studied extragalactic survey regions.
Surveys such as the Red-sequence Cluster Survey and its successor RCS-2 \citep{gladders_red-sequence_2005, gilbank_rcs2_2011}, the Spitzer Adaptation of the Red-sequence Cluster Survey \citep[SpARCS;][]{wilson_sparcs_2009, muzzin2009}, the
ESO Distant Cluster Survey \citep[EDisCS;][]{white_ediscs_2005}, and, more recently, C3VO \citep{hung_c3vo_2025} have exploited the rich multiwavelength data in such fields to cross-validate detection techniques and characterize cluster physical properties. Most recently, JWST programs such as SLICE \citep[PI: Mahler;][]{cerny_slice_2026} and VENUS \citep[PID~6882, PIs: Fujimoto \& Coe;][]{fujimoto_venus_2025} are providing high-resolution characterizations of cluster mass distributions and member galaxy populations. This ancillary data allows detailed derivation of cluster properties, thereby providing a benchmark for interpreting wide-area catalogs.

In this paper, we focus on the Bo\"otes field: a 9.3 deg$^2$ region near the North Galactic Pole that was originally targeted by the NOAO Deep Wide-Field Survey \citep[NDWFS,][]{ndwfs_1999}. This field has since been extensively observed across the electromagnetic spectrum and was used by the ISCS to identify 335 galaxy clusters, including 106 at $z > 1$, based on data from the IRAC Shallow Survey \citep{eisenhardt_infrared_2004}. ISCS clusters have been the focus of a number of follow-up studies that examined the properties of galaxies in high-redshift environments. These include investigations of star formation in galaxy clusters \citep{brodwin_2013,alberts_2014,alberts_2016,alberts_2022}, the effects of dust obscuration on star formation \citep{alberts_2021,mckinney_2022}, brightest cluster galaxy evolution \citep{lin_2013}, AGN activity and black hole growth \citep{martini_2013,alberts_2016}, emission-line spectroscopy of star-forming members \citep{zeimann_2013}, and quenching in cluster galaxies \citep{wagner_2017,sazonova_2020,samanso_2025}.

The IRAC Shallow Survey was extended by the deeper Spitzer Deep Wide-Field Survey \citep[SDWFS,][]{ashby2009} in the Bo\"otes field. This enabled the IRAC Distant Cluster Survey (IDCS), leading to the discovery of some of the most distant spectroscopically confirmed galaxy clusters known at the time. Such discoveries include the $z=1.89$ cluster IDCS J1433.2+3306 \citep{zeimann_2012}, and the $z=1.75$ cluster IDCS J1426.5+3508 \citep{brodwin_2016}. The latter has been extensively studied and remains among the most massive clusters known at $z > 1.5$ \citep{Kim_2025}. It has mass estimates from X-ray \citep{stanford_2012,brodwin_2016}, SZ \citep{brodwin_2012}, and weak-lensing \citep{mo_2016} observations, as well as strong lensing features \citep{gonzalez_2012}. ISCS and IDCS clusters have also enabled studies of galaxy evolution in dense environments at $z > 1$, including stellar structure \citep{andreon_2016}, molecular gas content \citep{alberts_2022}, and supernova cosmology \citep{hayden_2021}.

However, neither the full ISCS catalog nor the full IDCS catalog which supplanted it have been published. In this work, we present a new, comprehensive IDCS catalog based on a reanalysis of the full SDWFS data set. This release differs from what was used in those original discoveries in two ways. First, the photometric redshifts used in this search are superior to those which enabled the original IDCS search due to the inclusion of several new photometric bands in the optical and near-infrared (NIR). Second, the detection code used for this search, \texttt{PZWav} \citep{pzwav}, is a modernized version of the original detection code used in \cite{iscs_paper}. \texttt{PZWav} implements a more robust merging of detections in different redshift slices, uses random sky maps to compute the significance of detections, and is significantly faster than the original code. The input parameters for the current search, such as the smoothing kernels, have also been optimized relative to the original version. \texttt{PZWav} was used to find groups and clusters in the Stripe 82 region using the Southern Photometric Local Universe Survey \citep[S-PLUS,][]{werner_s-plus_2023}, and is the basis of the MaDCoWS2 \citep{thongkham_massive_2024_b,thongkham_massive_2024_a}. \texttt{PZWav} is also one of the two galaxy cluster identification codes selected for use by the Euclid Consortium \citep{euclid_pzwav}.

Using the first data release (DR1) of DESI, we identify member galaxies for 38 clusters with at least 10 spectroscopic members each, enabling measurements of cluster redshifts and velocity dispersions using both biweight and gapper estimators (\S \ref{sec:desi}). Dynamical masses are then inferred from the velocity dispersions using the scaling relation of \citet{munari_relation_2013}, yielding a median mass of $1.1 \times 10^{14} \, \mathrm{M}_\odot$ for this subset. Of these 38 clusters, 23 represent first-time spectroscopic confirmations, while 15 have previous literature identifications.

This paper is organized as follows. \S \ref{sec:data} describes our data, and how the photometric redshifts were estimated. \S \ref{sec:algorithm} explains the cluster search code. \S \ref{sec:idcs} presents the IDCS cluster candidate catalog, including the accuracy of our photometric redshifts. We also present redshift, velocity dispersion and mass estimates for a subset of the catalog, a comparison between the IDCS catalog and other published results, and a characterization of the survey completeness and purity from simulations. We present our conclusions in \S \ref{sec:conc}. We assume a flat $\Lambda$CDM cosmology from \cite{planck20} with $H_0 = 67.4 \rm{\; km \; s^{-1} Mpc^{-1}}$ and $\Omega_m = 0.315$, and magnitudes in the Vega system.

\section{Survey Data} \label{sec:data}

Within the NDWFS Boötes region, we make use of the photometric catalog of \cite{duncan_photometric_2018}, which is an updated and extended version of the catalog first presented in \cite{brown_evolving_2007}, selected at 4.5 $\mu$m. It includes optical data from NDWFS \citep{ndwfs_1999}; infrared (IR) data from the SDWFS \citep{ashby2009}; NIR data from the Infrared Boötes Imaging Survey \citep[IBIS,][]{IBIS:2010}; and \textit{u}-band and \textit{y}-band data from the Large Binocular Cameras \citep[LBC,][]{LBT:2013}. Table \ref{tab:survey_charact} shows the filters for each of these surveys, their 5$\sigma$ limits, and the original survey references. The depth columns distinguish between the shallower IRAC data used for the original ISCS \citep{iscs_paper}, which was limited to the 3.6 and 4.5~$\mu$m bands, and the deeper, more extensive photometry used in the present IDCS catalog.

For consistency across systems, we convert the $5 \sigma$ LBC \textit{u/y} depths of 25.2/21.6 AB magnitudes to Vega magnitudes using AB–Vega offsets of 0.9/0.6 mag based on the calibration by \citet{lbt_instruments_2008} for \textit{u} and the value given by the Gemini Observatory\footnote{https://www.gemini.edu/observing/resources/magnitudes-and-fluxes} for \textit{y}, yielding Vega magnitudes of 24.3/21 in these bands. The quoted 5$\sigma$ depths are taken directly from the original survey publications listed in Table~\ref{tab:survey_charact}. These values follow the depth definitions used by each survey, and we do not re-derive depths from the imaging.

\begin{deluxetable*}{lllcc}
\tablecaption{Survey characteristics. The 5$\sigma$ depth columns list the limiting magnitudes (Vega) for the ISCS and IDCS catalogs, respectively.\label{tab:survey_charact}}
\tablewidth{800pt}
\tablehead{\colhead{Survey} & \colhead{Band} & \colhead{Wavelength} & \multicolumn{2}{c}{5$\sigma$ limit} \\
& & \colhead{($\mu$m)} & \colhead{ISCS} & \colhead{IDCS}}
\startdata
  NDWFS & $B_W$ & 0.45 & 27.1$^{1}$ & 27.1$^{1}$ \\
        & $R$ & 0.65 & 26.1$^{1}$ & 26.1$^{1}$ \\
        & $I$ & 0.8 & 25.4$^{1}$ & 25.4$^{1}$ \\
  IRAC & [3.6] & 3.6 & 19.1$^{2}$ & 19.77$^{3}$ \\
       & [4.5] & 4.5 & 18.3$^{2}$ & 18.83$^{3}$ \\
       & [5.8] & 5.8 & -- & 16.50$^{3}$ \\
       & [8.0] & 8.0 & -- & 15.82$^{3}$ \\
  IBIS & $J$ & 1.2 & -- & 20.5$^{4}$ \\
       & $H$ & 1.65 & -- & 19.7$^{4}$ \\
       & $K_S$ & 2.15 & -- & 19.0$^{4}$ \\
  LBC & $u$ & 0.35 & -- & 24.3$^{5}$ \\
      & $y$ & 0.98 & -- & 21$^{6}$
\enddata
\tablenotetext{}{References: [1]~\citet{iscs_paper}; [2]~\citet{eisenhardt_infrared_2004}; [3]~\citet{ashby2009}; [4]~\citet{IBIS:2010}; [5]~\citet{LBT:2013}; [6]~\citet{chung_2014}.}
\end{deluxetable*}

We compute photometric redshifts based on an empirical template-fitting code developed by \cite{brodwin2006}. This same technique was used in \cite{iscs_paper} for the ISCS, which used data from the IRAC Shallow Survey (3.6 and $4.5 \; \mu$m filters, with depths of 90 seconds per pixel) and the NDWFS (\textit{Bw}, \textit{R}, and \textit{I} filters).
The IDCS presented here improves upon this by adding the IBIS and LBC imaging, and by using the deeper IRAC data from SDWFS \citep{ashby2009}, with exposures of 360 seconds or more per pixel. This added depth helps extend to higher redshifts and enables the 5.8 and $8.0 \; \mu$m IRAC band photometry to provide useful constraints on photometric redshifts, whereas those wavelengths were omitted in the ISCS.

We adopt a combination of the empirically-derived model templates Ell5, Ell13, S0, Sa, Sb, Sc, Sd, Spi4, and M82 from \cite{Polletta:2007}, and the \cite{Coleman:1980} elliptical template. These templates span the rest-frame wavelengths of $\sim 0.1$---$8 \; \mu$m probed by our filters for SDWFS galaxies at $0 < z < 2$. Extensions to NIR and far-ultraviolet (FUV) were made using \cite{bruzual_stellar_2003} models. Additional improvements in photometric redshift accuracy may be possible by selecting from multiple templates for each source (e.g., \citealt{duncan_photometric_2018}), but we do not explore this in the present work.

\begin{figure}[ht]
    \centering
    \includegraphics[width=\columnwidth]{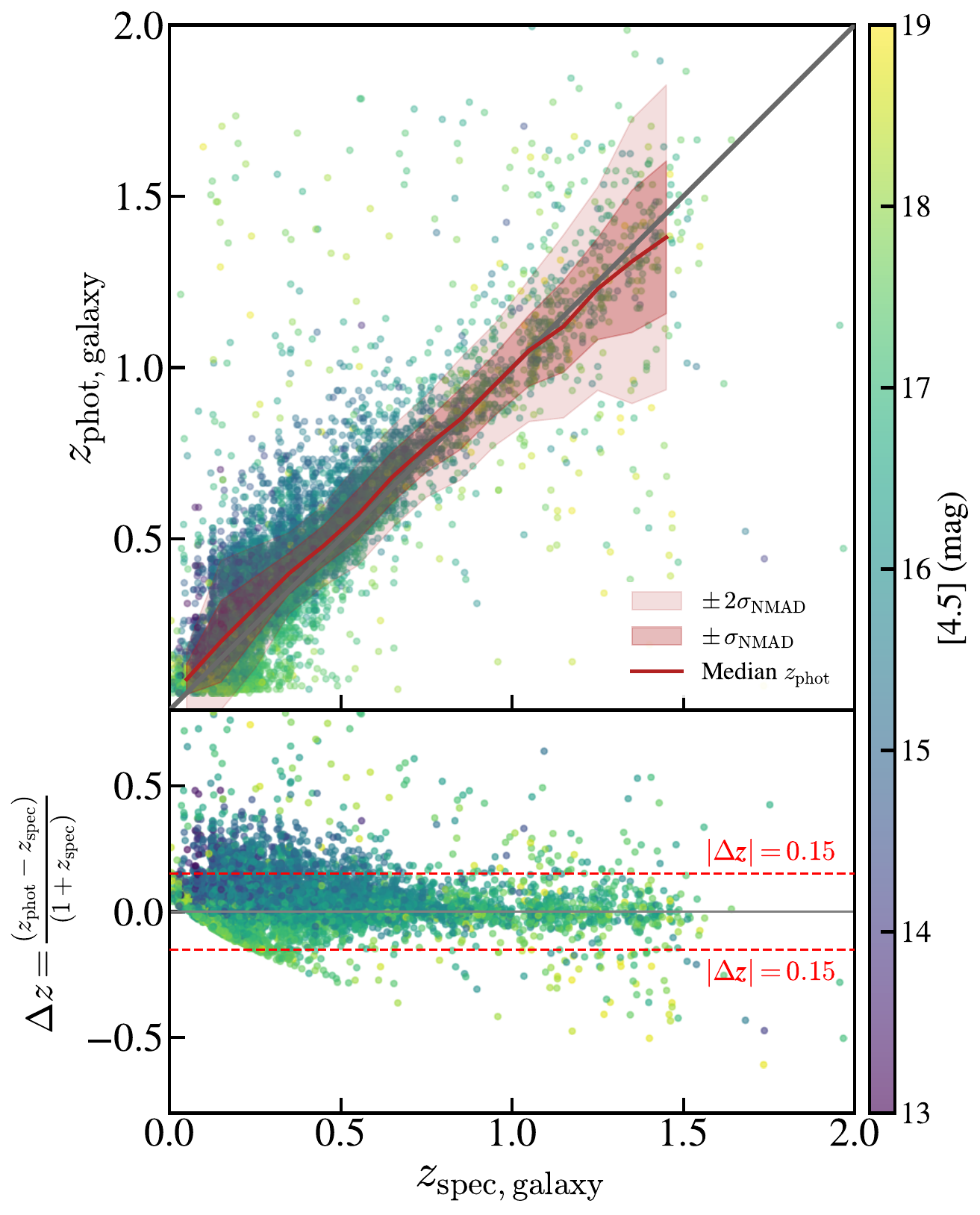}
    \caption{Photometric-redshift performance for galaxies with secure spectroscopic redshifts in the IDCS field. Top: photometric versus spectroscopic redshift, color-coded by $[4.5]$ magnitude. The solid red line is the median $z_{\rm phot}$ in bins of $z_{\rm spec}$, and the shaded regions show the robust scatter about the median, $\pm\sigma_{\rm NMAD}$ (inner) and $\pm2\sigma_{\rm NMAD}$ (outer), computed in bins containing at least 50 galaxies; the gray solid line is the one-to-one relation. The median and scatter bands are shown for $z_{\rm spec} < 1.5$, beyond which the spectroscopic sample becomes too sparse to be representative. Bottom: normalized residual $\Delta z = (z_{\rm phot}-z_{\rm spec})/(1+z_{\rm spec})$ as a function of spectroscopic redshift, color-coded by $[4.5]$ magnitude; horizontal lines mark $\Delta z = 0$ and the catastrophic-outlier threshold $|\Delta z| = 0.15$.}
    \label{fig:photozplot}
\end{figure}

Figure \ref{fig:photozplot} compares our photometric redshift estimates to spectroscopic data from \citet{stanford_2012}, \citet{Kochanek:2012}, \citet{zeimann_2013}, and \citet{brodwin_2013}. The upper panel shows $z_{\rm phot}$ as a function of $z_{\rm spec}$, color-coded by $[4.5]$ magnitude, with the median relation and its robust scatter ($\pm\sigma_{\rm NMAD}$, $\pm2\sigma_{\rm NMAD}$) overlaid in bins of $z_{\rm spec}$; the solid line shows the one-to-one relation. The lower panel shows $\Delta z$ as a function of $z_{\rm spec}$, color-coded by $[4.5]$ magnitude. The scatter is small and the median tracks the one-to-one relation closely out to $z \sim 1.5$, with the scatter broadening toward higher redshift.

Another improvement with respect to \cite{iscs_paper} is that we are no longer using a prior derived from a preliminary (no-prior) photometric redshift run. This prior helped the photometric redshifts not break down out to $z \sim 1.5$, but imposed a bias that had to be corrected using a fitting function. This time with deeper photometry and more bands there is no need to apply such a prior. We still observe a slight bias in Figure \ref{fig:photozplot} where photometric redshifts tend to be overestimated, particularly at $z \lesssim 0.8$.

\subsection{Photometric Redshift Accuracy}
\label{sec:photoz_accuracy}

Here we characterize the quality of the galaxy photometric redshifts that serve as input to the cluster detection. We restrict the comparison to objects spectroscopically classified as galaxies with secure redshifts (quality classes A and B), excluding stars and spectroscopically identified active galactic nuclei (AGN), whose spectral energy distributions are not well represented by the galaxy templates used in the photometric-redshift fitting. This yields $12{,}612$ galaxies with spectroscopic redshifts spanning $0 < z \leq 2$.

We quantify the accuracy using the normalized residual $\Delta z$. We report three standard metrics: the bias, defined as the median of $\Delta z$; the robust scatter $\sigma_{\rm NMAD} = 1.48 \times {\rm median}(|\Delta z - {\rm median}(\Delta z)|)$; and the catastrophic outlier fraction $f_{\rm out}$, defined as the fraction of galaxies with $|\Delta z| > 0.15$. For the full galaxy sample we find a bias of $+0.030$, $\sigma_{\rm NMAD} = 0.057$, and $f_{\rm out} = 9.3\%$. The small positive bias is consistent with the mild tendency for photometric redshifts to be overestimated at $z \lesssim 0.8$ noted in the cluster-level comparison (\S\ref{sec:desi}, Equation~\ref{eq1}). The galaxy-level scatter is a factor of $\sim\!3$ larger than the cluster-level value of $\sigma_z/(1+z) = 0.017$, as expected since cluster redshifts average over many member galaxies.

Table~\ref{tab:photoz} presents these metrics in bins of spectroscopic redshift and $[4.5]$ magnitude, which traces the galaxy stellar mass at $0.7 < z < 2$ \citep{iscs_paper}, so the magnitude bins serve as a proxy for stellar-mass dependence at these redshifts. The performance is good below $z \sim 1$, and best at $0.5 < z < 1$, where the scatter and outlier fraction reach their lowest values. The robust scatter remains small out to $z \sim 1.5$, though the catastrophic outlier fraction rises substantially to $18.9\%$ over this interval. The scatter increases markedly only beyond $z \sim 1.5$, where the spectroscopic sample also becomes sparse (only $19$ galaxies at $1.5 < z < 2.0$) and the metrics are correspondingly uncertain.

The dependence on apparent magnitude is less straightforward. The scatter $\sigma_{\rm NMAD}$ is not monotonic with magnitude: it is $0.052$ at $[4.5] < 16$, drops to a minimum of $0.045$ at $16 < [4.5] < 17$ as the lowest-redshift ($z < 0.2$) galaxies fall out of the sample, and then rises to $0.108$ at $[4.5] > 18$ as faint galaxies with low signal-to-noise photometry dominate. The catastrophic outlier fraction is lowest at intermediate magnitudes ($4.9\%$ for $16 < [4.5] < 17$) and increases at both the bright and faint ends. For the faintest galaxies, the outlier fraction reaches $32.2\%$ at $[4.5] > 18$, and most of these outliers have underestimated redshifts ($z_{\rm phot} \ll z_{\rm spec}$), as expected when the photometry has low signal-to-noise. At the bright end, the outlier fraction is $11.4\%$ for $[4.5] < 16$. These outliers are mostly low-redshift galaxies with slightly overestimated photometric redshifts, consistent with the small positive bias discussed above. The degradation toward higher redshift likely reflects not only the fainter magnitudes of distant galaxies but also the changing galaxy population, as the fraction of blue, star-forming galaxies with less well-constrained photometric redshifts increases with redshift. Both effects contribute to the decrease in detection completeness quantified in \S\ref{sec:completeness}, and together they indicate that the survey characterization becomes less certain beyond $z \sim 1.5$.

\begin{deluxetable}{lrrrr}
\tabletypesize{\footnotesize}
\tablecaption{Galaxy photometric-redshift accuracy\label{tab:photoz}}
\tablehead{
  \colhead{Bin} & \colhead{$N$} & \colhead{Bias} &
  \colhead{$\sigma_{\rm NMAD}$} & \colhead{$f_{\rm out}$}
}
\startdata
\cutinhead{By spectroscopic redshift}
$0.0<z<0.5$ & $10{,}096$ & $+0.035$ & $0.058$ & $9.3\%$ \\
$0.5<z<1.0$ & $1{,}972$  & $+0.016$ & $0.042$ & $6.1\%$ \\
$1.0<z<1.5$ & $525$      & $-0.007$ & $0.062$ & $18.9\%$ \\
$1.5<z<2.0$ & $19$       & $-0.078$ & $0.199$ & $52.6\%$ \\
\cutinhead{By $[4.5]$ magnitude}
$[4.5]<16$      & $6{,}104$ & $+0.054$ & $0.052$ & $11.4\%$ \\
$16<[4.5]<17$   & $4{,}632$ & $+0.015$ & $0.045$ & $4.9\%$ \\
$17<[4.5]<18$   & $1{,}646$ & $-0.037$ & $0.072$ & $10.3\%$ \\
$[4.5]>18$      & $230$     & $-0.018$ & $0.108$ & $32.2\%$ \\
\cutinhead{Full sample}
All & $12{,}612$ & $+0.030$ & $0.057$ & $9.3\%$ \\
\enddata
\end{deluxetable}

\section{Cluster Detection} \label{sec:algorithm}

The \texttt{PZWav} code \citep{pzwav} searches for overdensities on fixed physical scales using a difference-of-Gaussians kernel. \texttt{PZWav} takes as input a catalog of galaxies with their coordinates and the photometric redshift probability density functions (PDFs; \S \ref{sec:data}). Galaxies are assigned to a data cube with a spatial pixel scale of $12\arcsec$ and redshift slices of half-width $\Delta z = 0.06$ (full width $0.12$) with centers spaced every $0.06$, so that adjacent slices overlap by $50\%$. Each galaxy contributes to every slice it overlaps, weighted by its PDF.

We apply a mask to minimize spurious detections associated with bright stars. The mask is constructed using the NDWFS third data release\footnote{https://noirlab.edu/science/data-services/other/ndwfs/data-access/dr3-descr} (DR3) mask files. We extend the sizes of those regions around masked stars and near the edges by 10$^{\prime\prime}$ in radius to effectively mask residual artifacts. After applying the mask to remove regions with poor photometry or high stellar density, the effective survey area is 7.25 deg$^2$.

We next apply a difference-of-Gaussians smoothing kernel to the data cubes, with the inner and outer Gaussians having $\sigma = 400$~kpc and $\sigma = 2$~Mpc, respectively, at the midpoint of each redshift slice, following \cite{thongkham_massive_2024_b, thongkham_massive_2024_a}. We also construct an alternate density map to estimate the noise associated with the detection process. In this map, each galaxy retains its redshift probability distribution but is assigned a new position drawn at random within the survey footprint (respecting the mask). This randomization preserves the overall redshift distribution of the galaxy sample while destroying the two-dimensional spatial structure, so that no real overdensities survive. The fluctuations of the peaks recovered from this randomized map provide the noise against which detections are measured. The signal-to-noise ratio (S\slash N) of each detection is determined from the amplitude of the peak in the smoothed density map together with the fluctuations in the randomized maps, assuming a Poisson background-noise distribution.

The code identifies the most significant peaks within each redshift slice and merges peaks separated by less than $1.5\,$Mpc and $\Delta z = 0.12$, preventing the redundant counting of the same structure detected in overlapping slices as well as of physically merging clusters. The output catalog includes for each detection equatorial coordinates, an estimated cluster redshift and a signal-to-noise ratio, computed as described above.

\section{The IDCS catalog} \label{sec:idcs}

\begin{figure}
\centering
\includegraphics[width=0.95\columnwidth]{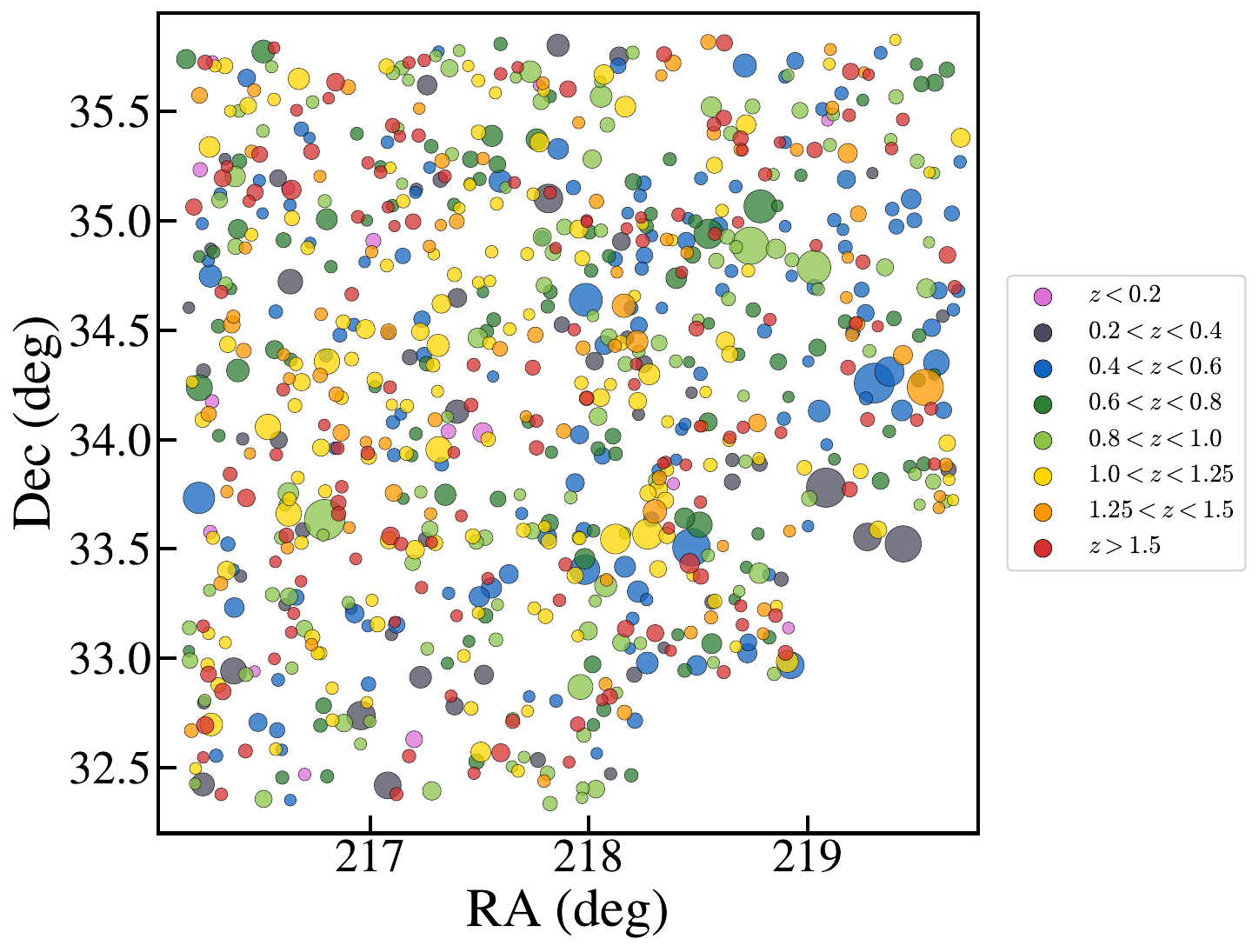}
\caption{Sky distribution of the 825 IDCS cluster candidates in the Bo\"{o}tes field, color-coded by photometric redshift. Point sizes scale with detection signal-to-noise ratio. The survey covers an effective area of 7.25~deg$^2$ after masking.}
\label{fig:sky_map}
\end{figure}

\begin{figure}[ht]
\includegraphics[width=0.95\columnwidth]{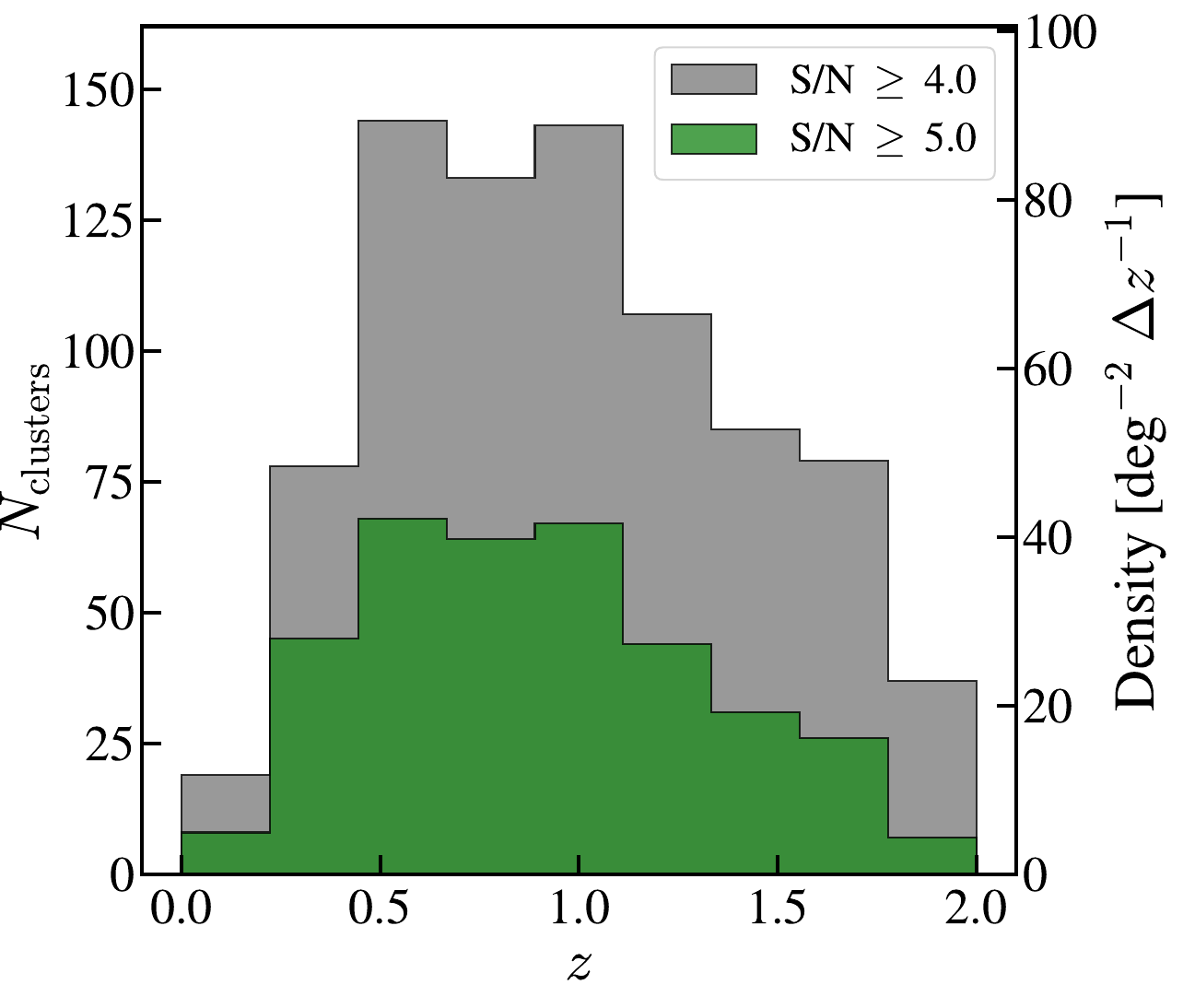}
\caption{The distribution of the IDCS cluster catalog as a function of redshift for two different signal-to-noise ratios. The right axis shows the cluster density assuming an effective survey area of 7.25 deg$^2$ after masking.}
\label{fig:histograms1}
\end{figure}

\begin{figure}[ht]
\includegraphics[width=0.95\columnwidth]{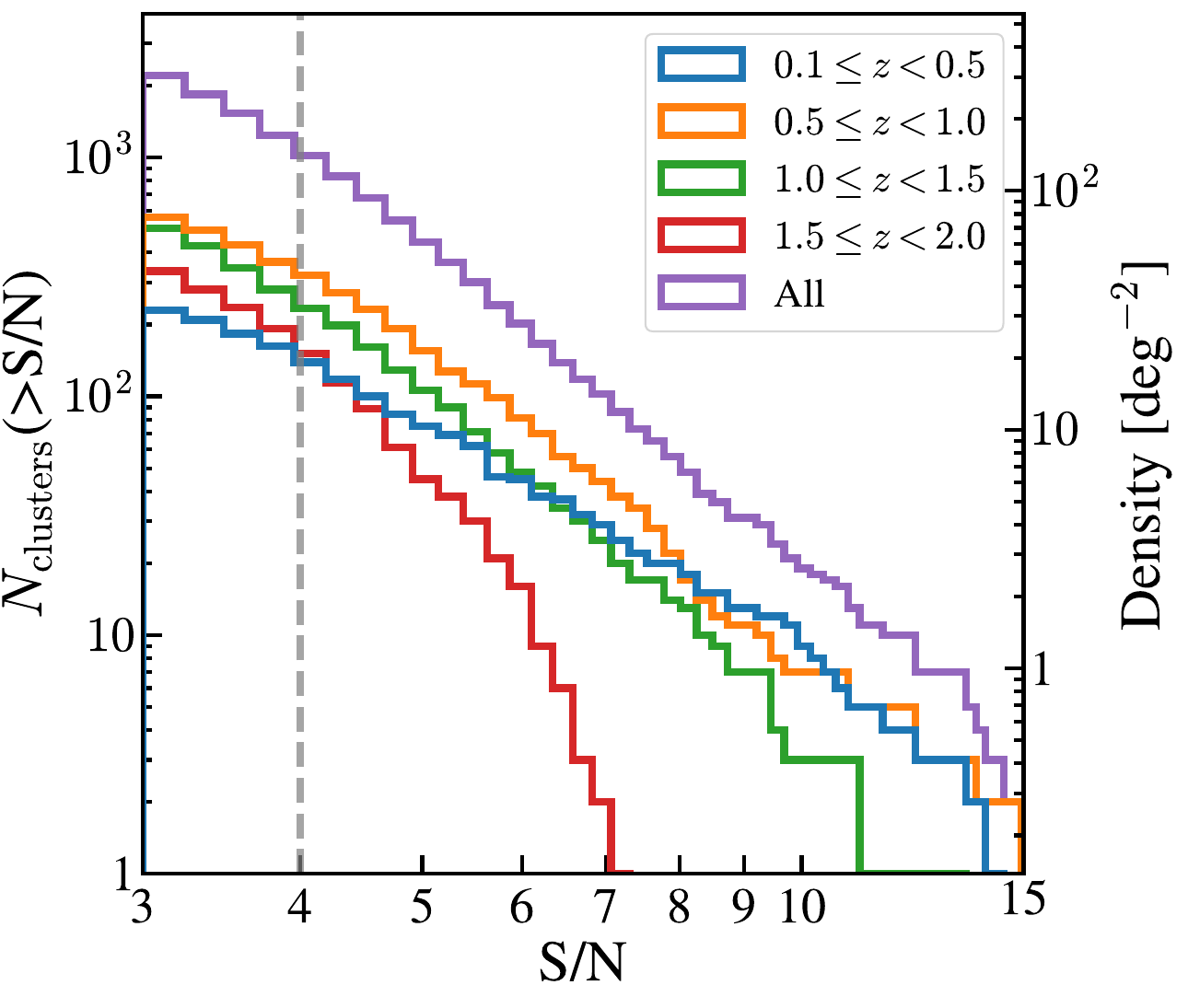}
\caption{Cumulative distribution of the catalog as a function of signal-to-noise ratio for five redshift intervals and the full sample. The right axis shows the cluster density assuming an effective survey area of 7.25 deg$^2$ after masking. The gray dashed line indicates the S\slash N = 4 threshold used in IDCS. Of the 825 clusters detected at S\slash N~$\geq$~4, 147 lie at $z \geq 1.5$.}
\label{fig:histograms2}
\end{figure}

Using the \texttt{PZWav} code, we detect overdensities down to S\slash N $=3$. For the published catalog we adopt a threshold of S\slash N $\geq 4$ to ensure higher purity, and limit the search to $z < 2$ due to a paucity of spectroscopic redshift calibrators beyond this range. This selection yields a final sample of 825 galaxy group and cluster candidates at $0.08 \le z < 2$. Figure~\ref{fig:sky_map} shows their spatial distribution, color-coded by redshift bin and sized by S\slash N. IDCS includes the well-known high-redshift cluster IDCS J1426.5+3508 (here referred to as IDCS J142633+350831), which has been spectroscopically confirmed and extensively characterized in previous studies \citep{stanford_2012, brodwin_2012, brodwin_2016, gonzalez_2012, mo_2016}, as well as IDCS J1433.2+3306 (here referred to as IDCS J143313+330654) at $z=1.89$ \citep{zeimann_2012}, which remained the most distant known galaxy clusters for more than a decade.

Table \ref{tab:cluster_catalog} lists IDCS's 30 highest S\slash N galaxy cluster candidates --- the full catalog is published in machine-readable format. Of the clusters and groups in the catalog, 247 match detections in the ISCS catalog, and 57 correspond to systems previously published in the literature, for which we provide cross-identifications in Table~\ref{tab:cluster_catalog} and discuss in greater detail in \S \ref{sec:comparison}. The photometric redshifts, $z_{\rm phot}$, listed in the table are outputs from \texttt{PZWav}. The associated uncertainty is empirically determined to be $\sigma_z / (1 + z) = 0.017$, as calculated in \S \ref{sec:comparison}. The redshift values labeled $z_{\rm DESI}$ are estimated from DESI DR1 \citep{desiDR1}, and $N_{\rm gals,DESI}$ indicates the number of galaxies within each group or cluster that have DESI spectra, as described in \S \ref{sec:desi}.

\begin{deluxetable*}{lcccccclcc}
\vspace{7mm}
\tablecaption{IDCS galaxy cluster candidates. The full catalog is available in machine-readable format; a portion is shown here. Column (1) provides the IAU designation for each cluster, based on the J2000.0 coordinates of the detection given in columns (2) and (3). Column (4) provides the signal-to-noise ratio, and Column (5) shows the photometric redshift for each detection. As described in \S \ref{sec:comparison}, the estimated redshift uncertainties are \( \sigma_z / (1 + z) = 0.017 \). Column (6) shows the spectroscopic redshifts calculated for the clusters with 10 or more galaxy members with DESI spectra, and column (7) the number of confirmed members (see \S \ref{sec:desi} for details). Columns (8), (9), and (10) list, respectively: the names of matching clusters in external catalogs; the projected distances between their positions and those in IDCS; and available spectroscopic redshifts from the literature, along with superscripted reference numbers. The references are: [1] \citet{vajgel_x-ray-selected_2014}; [2] \citet{rykoff_redmapper_2016}; [3] \citet{gonzalez_massive_2019}; [4] \citet{brodwin_2011}; [5] \citet{iscs_paper}; [6] \citet{elston2006}; [7] \citet{koulouridis2021}; [8] \citet{rykoff_redmapper_2014}; [9] \citet{snyder2012}; [10]~\citet{takey_2xmmisdss_2014}; [11]~\citet{brodwin_2013}; [12]~\citet{zeimann_2013}; [13]~\citet{stanford_2012}; [14]~\citet{zeimann_2012}; [15]~\citet{abdullah2020}; [16]~\citet{proctor2015}; [17]~\citet{mcconnachie2009}; [18]~\citet{duplancic2018}; [19]~\citet{zheng2020}.
\label{tab:cluster_catalog}}
\tablewidth{700pt}
\tabletypesize{\scriptsize}
\tablehead{
\colhead{ID} & \colhead{RA} & 
\colhead{DEC} & \colhead{S\slash N} & \colhead{$z_{\mathrm{phot}}$} & \colhead{$z_{\mathrm{DESI}}$} & \colhead{$N_{\mathrm{gals,DESI}}$} & \colhead{Literature ID} & \colhead{Distance} & \colhead{Literature $z_{\mathrm{spec}}$}  \\ 
\colhead{} & \colhead{(deg)} & \colhead{(deg)} & \colhead{} & 
\colhead{} & \colhead{} & \colhead{} & \colhead{} & \colhead{(arcsec)} & \colhead{}
 \\ 
\colhead{(1)} & \colhead{(2)} & \colhead{(3)} & \colhead{(4)} & 
\colhead{(5)} & \colhead{(6)} & \colhead{(7)} & \colhead{(8)} & \colhead{(9)} & \colhead{(10)}
}
\startdata
IDCS J142710+333801 & 216.791887 & 33.633524 & 14.9  & 0.86 & 0.85 & 16 &  &  &   \\
IDCS J143713+341529 & 219.304166 & 34.257992 & 14.7  & 0.53 & 0.55 & 11 & [VJL2014] XBS 38 & 31 & $0.54^{\href{https://ui.adsabs.harvard.edu/abs/2014ApJ...794...88V/abstract}{[1]}}$ \\
 &  &  &  &  &  &  & CXOXB J143714.3+341503 & 51 & $0.54^{\href{https://ui.adsabs.harvard.edu/abs/2016ApJS..224....1R/abstract}{[2]}}$ \\
IDCS J143620+334649 & 219.083925 & 33.780215 & 14.4  & 0.38 & 0.34 & 21 & [VJL2014] XBS 33 & 59 & $0.34^{\href{https://ui.adsabs.harvard.edu/abs/2014ApJ...794...88V/abstract}{[1]}}$ \\
 &  &  &  &  &  &  & [RRB2014] RM J143626.1+334657.5 & 73 & $0.34^{\href{https://ui.adsabs.harvard.edu/abs/2016ApJS..224....1R/abstract}{[2]}}$ \\
IDCS J143352+333028 & 218.468622 & 33.507724 & 13.8  & 0.44 &  &  & [RRB2014] RM J143356.6+333001.5 & 57 & $0.42^{\href{https://ui.adsabs.harvard.edu/abs/2016ApJS..224....1R/abstract}{[2]}}$ \\
IDCS J143457+345311 & 218.737421 & 34.886303 & 13.7  & 0.90 & 0.92 & 12 & MOO J1434+3453 & 27 & $0.92^{\href{https://ui.adsabs.harvard.edu/abs/2019ApJS..240...33G/abstract}{[3]}}$ \\
IDCS J143809+341419 & 219.537749 & 34.238523 & 13.3  & 1.38 &  &  & ISCS J143809+341419 & 4 & $1.41^{\href{https://ui.adsabs.harvard.edu/abs/2011ApJ...732...33B/abstract}{[4]}}$ \\
IDCS J143745+333123 & 219.436400 & 33.523089 & 13.3  & 0.30 &  &  & [SPD2011] 37360 & 37 & $0.22^{\href{https://ui.adsabs.harvard.edu/abs/2016ApJS..224....1R/abstract}{[2]}}$ \\
 &  &  &  &  &  &  & [VJL2014] XBS 42 & 38 & $0.22^{\href{https://ui.adsabs.harvard.edu/abs/2014ApJ...794...88V/abstract}{[1]}}$ \\
IDCS J143607+344715 & 219.028709 & 34.787413 & 12.3  & 0.91 & 0.93 & 19 &  &  &   \\
IDCS J143156+343820 & 217.985519 & 34.638860 & 12.1  & 0.40 &  &  & [VJL2014] XBS 20 & 15 & $0.35^{\href{https://ui.adsabs.harvard.edu/abs/2014ApJ...794...88V/abstract}{[1]}}$ \\
 &  &  &  &  &  &  & [RRB2014] RM J143155.3+343805.9 & 22 & $0.35^{\href{https://ui.adsabs.harvard.edu/abs/2016ApJS..224....1R/abstract}{[2]}}$ \\
IDCS J143508+350358 & 218.783963 & 35.066003 & 12.1  & 0.74 &  &  & [VJL2014] XBS 27 & 12 & $0.73^{\href{https://ui.adsabs.harvard.edu/abs/2014ApJ...794...88V/abstract}{[1]}}$ \\
IDCS J142452+334401 & 216.217291 & 33.733506 & 11.5  & 0.42 & 0.35 & 12 & [RRB2014] RM J142452.2+334401.0 & 1 & $0.36^{\href{https://ui.adsabs.harvard.edu/abs/2016ApJS..224....1R/abstract}{[2]}}$ \\
IDCS J143304+333406 & 218.268968 & 33.568423 & 11.1  & 1.04 &  &  & ISCS J1433.1+3334 & 29 & $1.07^{\href{https://ui.adsabs.harvard.edu/abs/2008ApJ...684..905E/abstract}{[5]}}$ \\
IDCS J143155+332420 & 217.980854 & 33.405522 & 10.9  & 0.50 &  &  &  &  &   \\
IDCS J143229+333243 & 218.120901 & 33.545397 & 10.9  & 1.11 &  &  & ClG J1432+3332 & 5 & $1.11^{\href{https://ui.adsabs.harvard.edu/abs/2006ApJ...639..816E/abstract}{[6]}}$ \\
 &  &  &  &  &  &  & ISCS J1432.4+3332 & 8 & $1.11^{\href{https://ui.adsabs.harvard.edu/abs/2008ApJ...684..905E/abstract}{[5]}}$ \\
IDCS J143730+341838 & 219.373632 & 34.310541 & 10.7  & 0.46 &  &  & [VJL2014] XBS 39 & 17 & $0.40^{\href{https://ui.adsabs.harvard.edu/abs/2014ApJ...794...88V/abstract}{[1]}}$ \\
 &  &  &  &  &  &  & XCLASS 1350 & 22 & $0.40^{\href{https://ui.adsabs.harvard.edu/abs/2021A&A...652A..12K/abstract}{[7]}}$ \\
IDCS J143411+345626 & 218.546829 & 34.940719 & 10.7  & 0.70 & 0.82 & 12 & [RRB2014] RM J143414.0+345629.8 & 34 & $0.57^{\href{https://ui.adsabs.harvard.edu/abs/2014ApJ...785..104R/abstract}{[8]}}$ \\
IDCS J143116+350608 & 217.814666 & 35.102151 & 10.5  & 0.22 & 0.19 & 77 & [KMA2007] 217.82750+35.10634 & 39 & $0.19^{\href{https://ui.adsabs.harvard.edu/abs/2016ApJS..224....1R/abstract}{[2]}}$ \\
 &  &  &  &  &  &  & [VJL2014] XBS 17 & 78 & $0.19^{\href{https://ui.adsabs.harvard.edu/abs/2014ApJ...794...88V/abstract}{[1]}}$ \\
IDCS J142750+324418 & 216.958172 & 32.738247 & 10.4  & 0.35 &  &  &  &  &   \\
IDCS J143706+333318 & 219.272937 & 33.554971 & 10.1  & 0.32 & 0.24 & 42 & [KMA2007] 219.27108+33.56086 & 26 & $0.24^{\href{https://ui.adsabs.harvard.edu/abs/2016ApJS..224....1R/abstract}{[2]}}$ \\
IDCS J142819+322509 & 217.080150 & 32.419130 & 9.8  & 0.31 &  &  &  &  &   \\
IDCS J143541+325806 & 218.922251 & 32.968244 & 9.8  & 0.42 &  &  & [RRB2014] RM J143541.2+325758.6 & 8 & $0.38^{\href{https://ui.adsabs.harvard.edu/abs/2016ApJS..224....1R/abstract}{[2]}}$ \\
IDCS J142530+325632 & 216.375952 & 32.942230 & 9.6  & 0.29 &  &  & [VJL2014] XBS  2 & 36 & $0.21^{\href{https://ui.adsabs.harvard.edu/abs/2014ApJ...794...88V/abstract}{[1]}}$ \\
IDCS J142452+341413 & 216.219154 & 34.236898 & 9.5  & 0.69 &  &  &  &  &   \\
IDCS J142915+335714 & 217.314038 & 33.953951 & 9.5  & 1.04 &  &  & ISCS J1429.2+3357 & 6 & $1.06^{\href{https://ui.adsabs.harvard.edu/abs/2012ApJ...756..114S/abstract}{[9]}}$ \\
IDCS J143821+342104 & 219.588351 & 34.351198 & 9.4  & 0.54 &  &  &  &  &   \\
IDCS J142608+340327 & 216.532663 & 34.057523 & 9.3  & 1.13 &  &  & ISCS J1426.1+3403 & 25 & $1.14^{\href{https://ui.adsabs.harvard.edu/abs/2012ApJ...756..114S/abstract}{[9]}}$ \\
IDCS J143401+333639 & 218.505280 & 33.610896 & 9.3  & 0.63 &  &  &  &  &   \\
IDCS J142630+333943 & 216.627292 & 33.661883 & 9.3  & 1.12 &  &  & ISCS J1426.5+3339 & 10 & $1.16^{\href{https://ui.adsabs.harvard.edu/abs/2012ApJ...756..114S/abstract}{[9]}}$ \\
IDCS J142712+342125 & 216.802262 & 34.356975 & 9.3  & 1.06 &  &  &  &  &   \\
IDCS J143150+325208 & 217.960755 & 32.868842 & 9.1  & 0.94 &  &  &  &  &   \\
\enddata
\end{deluxetable*}

Figure~\ref{fig:histograms1} shows the number of galaxy clusters (and their projected number densities) as a function of redshift for two different S\slash N cuts, out to $z=2$. The redshift distribution peaks around $z=1$, with median redshifts of $z_{\rm median} = 0.93$ for S\slash N $\geq 4$ and $z_{\rm median} = 0.86$ for S\slash N $\geq 5$. Of the 825 detected clusters, 147 lie at $z \geq 1.5$.

Four cluster candidates have final photometric redshifts at $z \leq 0.10$, including one high-significance detection (IDCS J143003+340153, S/N $= 7.3$) at $z=0.09$. We note that the PZWav search is performed over $z \geq 0.10$, while the final cluster redshifts are determined in a subsequent photometric-redshift refinement step. These low-redshift systems therefore originate from detections in the lowest search-redshift slices whose refined redshifts shift below the search limit. At $z=0.08$, the angular scale is 1.52~kpc/arcsec, so the 3-degree extent of the NDWFS field still spans a substantial physical scale ($\sim16$~Mpc), allowing such nearby systems to be identified within the survey area. These low-redshift systems may be underrepresented in spectroscopic follow-up surveys like DESI DR1, which tends to select against very nearby galaxies. Nonetheless, detections near the low-redshift boundary ($z \lesssim 0.14$), corresponding to the first \texttt{PZWav} redshift slice (centered at $z = 0.10$), are subject to larger photometric-redshift uncertainties and to edge effects in the density reconstruction, and their redshifts should therefore be treated with greater caution. For statistical analyses requiring a robustly characterized sample, we therefore recommend restricting the catalog to $z>0.14$.

Figure~\ref{fig:histograms2} displays the cumulative S\slash N distribution across different redshift intervals. The S\slash N distributions show broadly similar shapes across all redshift intervals, with the main difference that the maximum S\slash N attained is lower for $z > 1$, and the $z < 0.5$ distribution is noticeably flatter. This behavior reflects the varying detection completeness and photometric-redshift quality across redshift that we quantify in \S\ref{sec:photoz_accuracy} and \S\ref{sec:selecfunc}. At low redshift ($z < 0.5$), cluster member galaxies are relatively bright and are recovered with high completeness and small photometric-redshift scatter ($\sigma_{\rm NMAD} = 0.058$; \S\ref{sec:photoz_accuracy}), leading to a broad range of signal strengths and a flatter S\slash N distribution. At higher redshift, cluster galaxies become progressively fainter, an increasing fraction falls below our detection limits, and their photometric redshifts scatter more strongly; the completeness analysis of \S\ref{sec:completeness} shows that the survey remains complete for massive halos to $z \sim 1.5$ but that the recovered signal declines toward higher redshift. This reduces the available signal from cluster members and constrains the range of achievable S\slash N values, producing the steeper S\slash N distribution observed at $z > 1$.

\subsection{Catalog Selection Function}
\label{sec:selecfunc}
 
To estimate the completeness and purity of the catalog as a function of halo mass and redshift, we apply PZWav to a mock galaxy catalog for which the underlying halo population is known, following the approach adopted for MaDCoWS2 \citep{thongkham_massive_2024_b}.
 
\subsubsection{Mock catalog}
\label{sec:mock}
 
We use the ultra-wide light cones from \citet{Yung2023_sims}, a set of five independent $2\,\mathrm{deg}^2$ realizations constructed from dark matter halos in the Small MultiDark-Planck (SMDPL) simulation \citep{Klypin2016_sims} and populated with galaxies using the Santa Cruz semi-analytic model. Each galaxy in the light cone carries predicted observed-frame photometry and is associated with a host halo of known virial mass \citep[$M_{\rm vir}$;][]{BryanNorman1998}, which we treat as the ground-truth cluster population. The five realizations together span $11.55\,\mathrm{deg}^2$, comparable to the effective IDCS survey area.

To emulate the IDCS selection, we degrade the mock to match the depth and photometric-redshift properties of the observed galaxy catalog. Because the IDCS catalog is selected at $4.5\,\mu$m, we apply the corresponding flux limit to the mock IRAC $4.5\,\mu$m (channel~2) photometry, converting from the AB system used in the mock to Vega magnitudes using the IRAC $4.5\,\mu$m offset \citep[$3.26$~mag;][]{Reach2005_IRACcalibration}. After this selection, the five realizations contain a total of $1.13\times10^{6}$ galaxies. We further restrict the mock to $z<2.3$, corresponding to the redshift range populated by the IDCS galaxy catalog and extending slightly beyond the $z<2$ limit adopted for the published cluster catalog.
 
Because PZWav operates on photometric-redshift probability density functions (PDFs) rather than point estimates, each mock galaxy must be assigned a realistic redshift PDF. We adopt an empirical procedure: for each mock galaxy we identify its nearest neighbor in the space of $4.5\,\mu$m magnitude and redshift among the real IDCS galaxies, and assign that galaxy the observed redshift PDF of its match, recentered on the mock galaxy's true redshift. Because the two catalogs are matched in both magnitude and redshift, the inherited PDFs reproduce the width and shape of the true IDCS photometric-redshift uncertainties: the median PDF full width at half maximum in the mock agrees with that of the real catalog to within 10\% in every redshift bin. The resulting mock galaxy catalog and PDFs are written in the same format as the real input and processed through PZWav with identical parameters (\S\ref{sec:algorithm}). Because the mock fields are uncontaminated by bright stars, we do not apply the bright-star mask used for the real data; the mask instead enters the survey selection function as an independent areal-coverage factor (the effective area is 78\% of the nominal field; \S\ref{sec:algorithm}).
 
\subsubsection{Matching}
\label{sec:matching}

We run PZWav separately on each of the five realizations and cross-match the resulting detections against the input halo catalog, considering halos with $M_{\rm vir} > 10^{13}\,M_\odot$. A detection is eligible to be associated with a halo when the two lie within a projected physical separation of $1\,$Mpc and $|\Delta z| \leq 0.1\,(1+z)$. To ensure a one-to-one correspondence, we adopt a mutual matching scheme, in which a halo and a detection are paired only if the detection is the highest signal-to-noise detection within the halo's aperture and the halo is, in turn, the preferred (most massive) halo for that detection. In each iteration, any detection and halo that mutually prefer one another are matched and removed, and the procedure repeats on the remaining objects, since removing a pair can free others to match. It converges when no new mutual pairs are found, after three iterations for our catalog. This prevents a single halo from being counted against multiple detections, or a single detection from being credited to multiple halos. Completeness is defined as the fraction of halos of a given mass and redshift that are matched to a detection, and purity as the fraction of detections matched to a halo. Matching is performed independently within each realization, since the five light cones sample independent volumes.

\subsubsection{Completeness}
\label{sec:completeness}

\begin{figure*}
\centering
\includegraphics[width=0.9\textwidth]{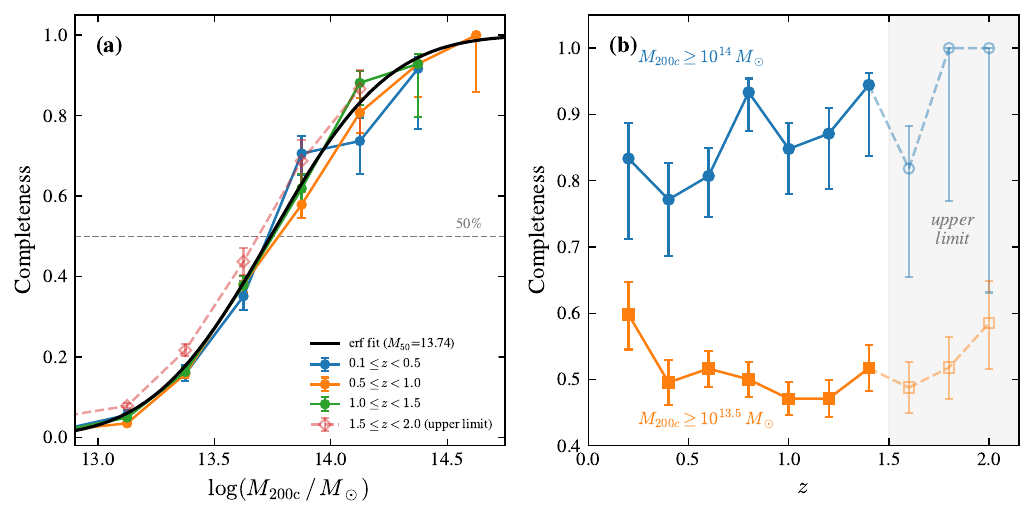}
\caption{Survey completeness from simulations. (a) Completeness as a function of halo mass ($M_{200c}$) for detections at S\slash N $\geq 4$, in four redshift bins (colors as in Figure~\ref{fig:histograms2}); the solid black curve is the best-fit error function, which reaches 50\% completeness at $\log(M_{200c}/M_\odot) = 13.74$, and the horizontal dashed line marks 50\% completeness. (b) Completeness as a function of redshift for cluster-scale halos ($M_{200c} \geq 10^{14}\,M_\odot$, circles) and for $M_{200c} \geq 10^{13.5}\,M_\odot$ (squares). In both panels, the $1.5 < z < 2.0$ bin is shown with open symbols and dashed lines, as it represents an upper limit: the real galaxy catalog becomes observationally incomplete at $z \gtrsim 1.5$. Error bars are Clopper--Pearson binomial confidence intervals.}
\label{fig:completeness}
\end{figure*}

Figure~\ref{fig:completeness}(a) shows the resulting completeness as a function of halo mass in four redshift bins, for detections at S\slash N $\geq 4$ (the threshold adopted for the published catalog), converting from $M_{\rm vir}$ to $M_{200c}$ using \textsc{colossus} \citep{Diemer2018_colossus} with the concentration model of \citet{Diemer2019_haloconcentrations}. The completeness follows the error function form expected for a Gaussian scatter between detectability and mass. We fit $f(\log M) = \tfrac{1}{2}\left[1 + \mathrm{erf}\!\left((\log M - \log M_{50})/\sqrt{2}\,\sigma\right)\right]$ to the measured completeness, obtaining a 50\% completeness mass of $\log(M_{50}/M_\odot) = 13.74 \pm 0.01$ with a width $\sigma = 0.40 \pm 0.01$ (in $M_{200c}$, for $z < 1.5$); the fit is stable whether we include all mass bins or only those above 30--40\% completeness. The completeness for $M_{200c} \geq 10^{14}\,M_\odot$ is $\sim\!86\%$ globally, ranging from 79\% at $z < 0.5$ to 89\% at $1 < z < 1.5$, and reaches 100\% for $M_{200c} \geq 10^{14.5}\,M_\odot$. Figure~\ref{fig:completeness}(b) shows the completeness for $M_{200c} \geq 10^{14}\,M_\odot$ as a function of redshift, confirming that cluster-scale halos are recovered with high and roughly uniform completeness out to $z \sim 1.5$; the best-fit $M_{50}$ varies by only $\sim\!0.06$~dex across redshift bins. This substantiates the interpretation in \S\ref{sec:idcs} that the redshift dependence of the S\slash N distribution (Figure~\ref{fig:histograms2}) reflects varying detection completeness with redshift.

For comparison with MaDCoWS2, whose 50\% completeness mass is $\log(M_{500c}/M_\odot) \approx 14.1$, we convert our result to the same mass definition and obtain $\log(M_{500c}/M_\odot) = 13.58 \pm 0.01$. The IDCS selection thus reaches $\sim0.5$~dex lower in mass than MaDCoWS2, consistent
with the greater depth of our IRAC selection ($[4.5] < 18.8$ Vega) relative to the WISE selection of MaDCoWS2.
 
We note three limitations of this analysis. First, the ultra-wide light cones contain few very massive halos owing to their limited volume, so the completeness at the highest masses ($M_{\rm vir} \gtrsim 10^{14.5}\,M_\odot$) is based on small numbers ($26$ halos across the five realizations) and carries correspondingly large uncertainty. Second, the mock galaxy population is volume-complete, whereas the real IDCS galaxy catalog becomes observationally incomplete at $z \gtrsim 1.5$: the surface density of real galaxies relative to the mock falls from near unity at $z<1.5$ to $\sim\!0.6$ at $1.5<z<2.0$ and $\sim\!0.1$ at $z>2$. The completeness measured here should therefore be regarded as reliable to $z \sim 1.5$ and as an upper limit at higher redshift. Both limitations affect the same high-mass, high-redshift regime and are consistent with the behavior noted for the comparable analysis in \citet{thongkham_massive_2024_b}.

A third limitation is that the redshift PDFs assigned to the mock galaxies depend only on magnitude and redshift, not on spectral type. In practice, red, early-type galaxies (which have a strong $4000$\,\AA\ break and therefore more accurate photometric redshifts) dominate cluster cores, whereas blue, star-forming galaxies with poorer photometric redshifts are more common in the field and become an increasing fraction of the population toward high redshift. Because our matching does not distinguish these populations, the mock assigns cluster galaxies a redshift PDF averaged over both types, which likely underestimates the photometric-redshift quality of the true cluster members at high redshift. The completeness we measure at $z \gtrsim 1.5$ should therefore be regarded as a conservative lower limit, as the early-type galaxies that drive cluster detection retain better photometric redshifts than the mock assumes.
 
\subsubsection{Purity}
\label{sec:purity}
 
\begin{figure}
\centering
\includegraphics[width=0.95\columnwidth]{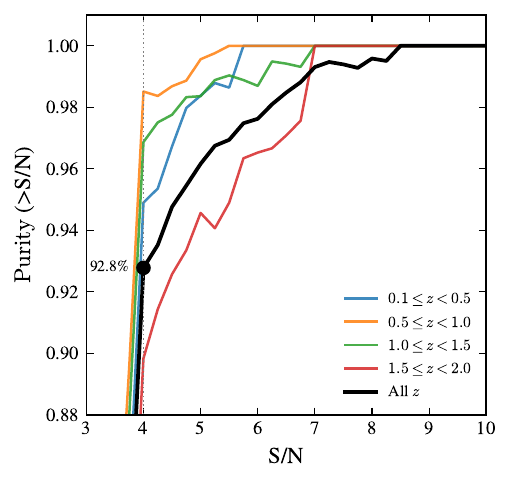}
\caption{Catalog purity as a function of S/N threshold, globally (black) and in four redshift bins (colors as in Figure~\ref{fig:histograms2}). The vertical dashed line marks the S\slash N $\geq 4$ threshold adopted for the published catalog, where the global purity is 93\%.}
\label{fig:purity}
\end{figure}

Figure~\ref{fig:purity} shows the catalog purity as a function of S/N threshold. At the adopted threshold of S\slash N $\geq 4$, the global purity is 93\%, rising to 96\% at S\slash N $\geq 5$. The purity remains high across the survey redshift range, ranging from 90\% to 99\% across redshift bins at $z < 2$. The high purity already at S\slash N $\geq 4$ supports the choice of this threshold for the published catalog (\S\ref{sec:idcs}). These values exceed the $>90\%$ purity reported for MaDCoWS2 and are consistent with expectations for cluster searches based on deep infrared imaging.

We note that the simulation-based purity estimate accounts for contamination from background fluctuations and projection effects represented in the mock catalog, but does not capture potential contamination associated with observational artifacts in the real imaging. As an empirical check, 17 of the 19 IDCS detections with S/N $>10$ have independent spectroscopic or X-ray confirmation, corresponding to a confirmation fraction of $89\%$. The remaining two systems are unconfirmed rather than known contaminants, so this provides a lower limit of $89\%$ on the purity of the high-S/N sample (and, equivalently, an upper limit of $11\%$ on its contamination based on the available confirmation data).

\subsubsection{Merging and Linking Lengths}
\label{sec:linking}

A potential concern with the peak-merging procedure is that nearby clusters may be incorrectly combined into a single detection, particularly within filaments or superclusters. We use the simulations to test the fixed linking lengths adopted for merging detections across redshift slices
(1.5~Mpc projected, $\Delta z = 0.12$; \S\ref{sec:algorithm}). Our mutual matching scheme pairs each halo and each detection at most once (\S\ref{sec:matching}), but a detection can still overlap the apertures of several halos, and a halo the apertures of several detections, before the one-to-one pairing is imposed. We quantify over-merging and fragmentation from these aperture overlaps.

Over-merging (a detection whose aperture contains two or more distinct halos with $M \geq 10^{14}\,M_\odot$) affects only 4 of 235 detections ($1.7\%$), and none contain three or more. Two of these are line-of-sight projections ($\sim\!300$--$400$~Mpc apart), one is a close pair consistent with an ongoing merger (1.5~Mpc), and one is a wider pair (5.7~Mpc) more likely tracing larger-scale structure. Conversely, fragmentation (a halo matched to more than one detection) affects only 3 of 236 halos ($1.3\%$), none of which is a clear case of a single cluster signal split into comparable detections. We conclude that the adopted linking lengths reliably keep distinct massive halos separate, with negligible over-merging or fragmentation at the cluster scale.

\subsection{Comparison with other catalogs} \label{sec:comparison}

We next compare our detections with other catalogs covering the NDWFS region, some of which provide additional spectroscopic confirmation or X-ray detections. These catalogs include the Northern Sky Optical Cluster Survey \citep{gal_northern_2003}, the AMF SDSS DR6 catalog \citep{szabo_optical_2011}, the MaxBCG catalog from SDSS \citep{KMA2007}, the Chandra X-Bo\"otes survey \citep{vajgel_x-ray-selected_2014}, the redMaPPer galaxy cluster catalog \citep{rykoff_redmapper_2016}, the Massive and Distant Clusters of WISE Survey \citep[MaDCoWS;][]{gonzalez_massive_2019}, and several catalogs containing more than 1.5 million clusters identified from photometric surveys \citep{wen_2009, wen_2011, wen_2017, wen_2021, WH2024}. MaDCoWS2 \citep{thongkham_massive_2024_b,thongkham_massive_2024_a}, is not included as it does not overlap the NDWFS field, being limited to dec $\leq$ +30 degrees. We define a match as occurring when the IDCS candidate and a cluster candidate from another catalog lie within 1 Mpc (projected physical separation at the cluster redshift) and have a redshift difference of $\Delta z \leq 0.2$.

\begin{figure}[ht]
\includegraphics[width=0.95\columnwidth]{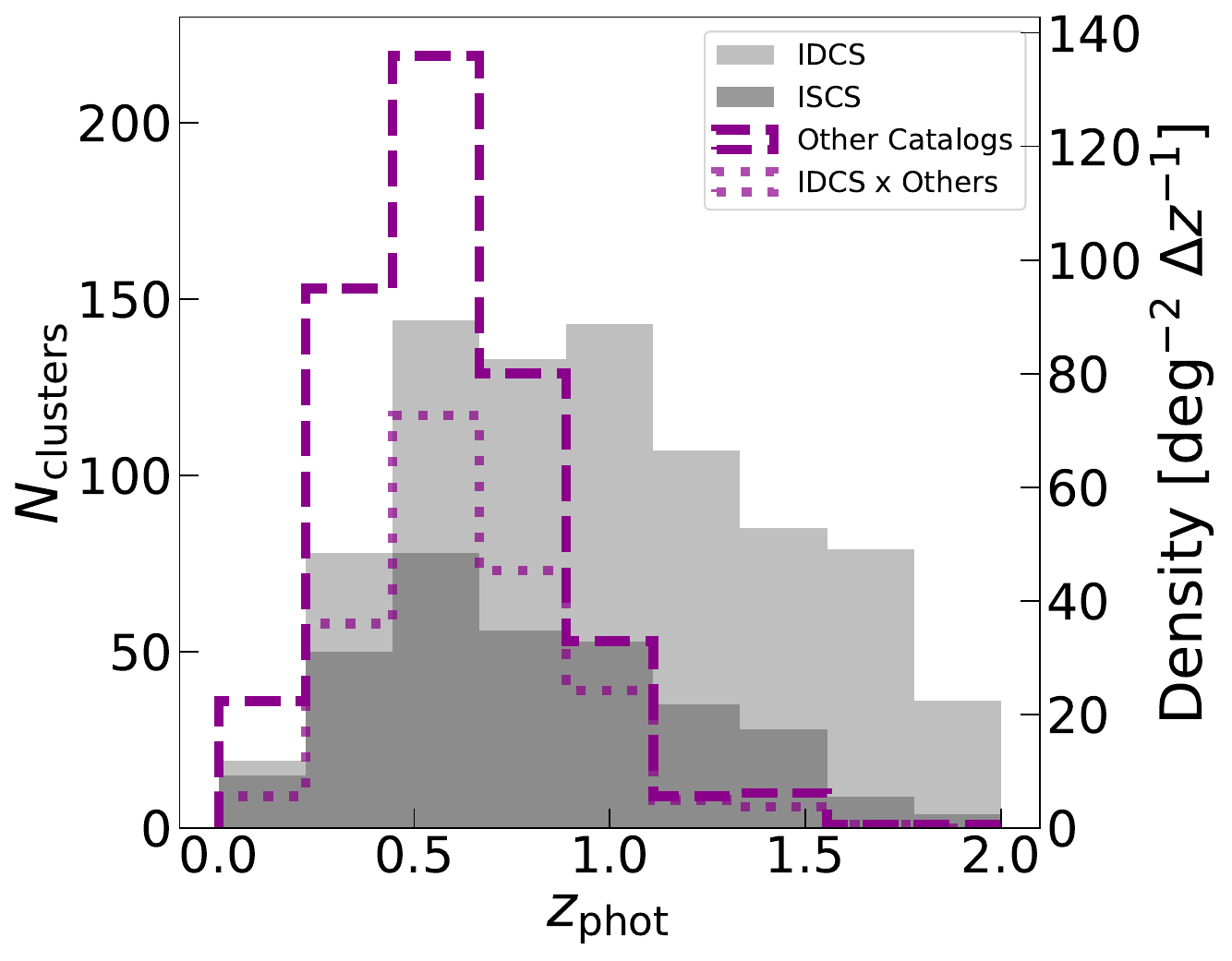}
\caption{Redshift distribution of galaxy clusters in the IDCS catalog (light gray) and its precursor, the ISCS catalog (dark gray). The redshift distribution for galaxy clusters from other photometric and spectroscopic surveys that lie within the NDWFS area is shown by the dashed purple line. Clusters from these external surveys that match IDCS detections are shown by the dotted purple line. IDCS recovers 95\% of high-significance (significance $> 1.3$) ISCS detections. The gap between the dashed and dotted lines at $z < 1$ reflects the complementary selection functions of infrared and optical surveys: optically-selected catalogs achieve higher completeness for lower-mass systems at $z \sim 0.5$--$0.8$, while IDCS uniquely recovers hundreds of clusters at $z > 1$ not identified in existing large-scale catalogs. Density calculations on the y-axis assume an effective survey area of 7.25 deg$^2$.}
\label{fig:histz_photoz}
\end{figure}

Figure~\ref{fig:histz_photoz} presents the redshift distribution of the 825 galaxy clusters in the IDCS catalog, alongside that of 565 galaxy clusters from the other surveys within the NDWFS area listed above. Among these, 311 clusters match IDCS detections under our spatial and redshift criteria (1 Mpc projected separation, $\Delta z \leq 0.2$). Also included is the distribution for ISCS (the precursor to IDCS described in \citealt{iscs_paper}), which contains 335 clusters.

IDCS recovers 247 of the 335 ISCS clusters (about $74\%$). To investigate the unrecovered systems, we examine the IDCS recovery fraction as a function of the ISCS detection significance $\sigma$, defined as the wavelet-detection significance determined from bootstrap resampling of galaxy positions and photometric-redshift PDFs \citep{iscs_paper} (Table~\ref{tab:iscs_recovery}). The recovery fraction increases with significance, from $64\%$ for $1.0 \leq \sigma < 1.2$ to $100\%$ at $\sigma \geq 2.0$. Above $\sigma = 1.5$, IDCS recovers $97\%$ ($28/29$) of ISCS systems, while $85\%$ ($75/88$) of the unrecovered systems lie in the lowest-significance bin. Consistent with this trend, the unmatched ISCS systems have a median significance of $1.1$. \citet{iscs_paper} estimated that $\sim 10\%$ of ISCS candidates may be spurious due to false positives near the detection threshold and projection effects, with contamination expected to be concentrated at low significance. Although the unrecovered fraction ($26\%$) is larger than this global contamination estimate, its strong concentration at low significance, together with the near-complete recovery of high-significance systems, suggests that a substantial fraction of the non-recoveries may be spurious ISCS detections rather than genuine clusters missed by IDCS.

The comparison with other surveys illustrates the different selection functions of infrared and optical cluster-finding methods. IDCS recovers $41\%$ of the optically selected clusters from \citet{WH2024}. The unmatched systems are predominantly lower-mass clusters at $z < 0.8$, with a median $M_{500} = 0.69 \times 10^{14}~M_\odot$, where optical photometric redshifts provide higher completeness. At $z > 1$, however, the IDCS catalog contains nearly 400 clusters, compared with only 36 in the external catalogs and 106 in ISCS. Thus, even compared with extensive catalogs such as the 1.58 million clusters identified from the DESI Legacy Imaging Surveys \citep{WH2024}, IDCS adds hundreds of detections in the Bo\"otes region that are not recovered by existing surveys. This highlights the advantage of deep, infrared-selected galaxy samples at high redshift, where IRAC is sensitive to the red, evolved galaxy populations in clusters.

\setlength{\tabcolsep}{12pt}
\global\deluxesidewaystablefalse
\centerwidetable
\begin{deluxetable*}{lrrr}
\tabletypesize{\small}
\tablecaption{IDCS recovery of ISCS systems by detection significance\label{tab:iscs_recovery}}
\tablehead{\colhead{ISCS significance} & \colhead{$N$} & \colhead{$N_{\rm rec}$} & \colhead{Recovered}}
\startdata
$1.0 \leq \sigma < 1.2$ & 210 & 135 & $64.3\%$ \\
$1.2 \leq \sigma < 1.5$ & 96  & 84  & $87.5\%$ \\
$1.5 \leq \sigma < 2.0$ & 21  & 20  & $95.2\%$ \\
$2.0 \leq \sigma < 2.5$ & 8   & 8   & $100.0\%$ \\
\hline
Total & 335 & 247 & $73.7\%$ \\
\enddata
\tablecomments{Fraction of ISCS systems \citep{iscs_paper} recovered by IDCS as a function of ISCS detection significance, using a matching criterion of $1$~Mpc projected separation and $|\Delta z| < 0.2$. Recovery is near-complete at high significance and declines toward the detection threshold, where ISCS contamination is expected to concentrate.}
\end{deluxetable*}
\setlength{\tabcolsep}{6pt}

\begin{figure}[ht]
\includegraphics[width=0.95\columnwidth]{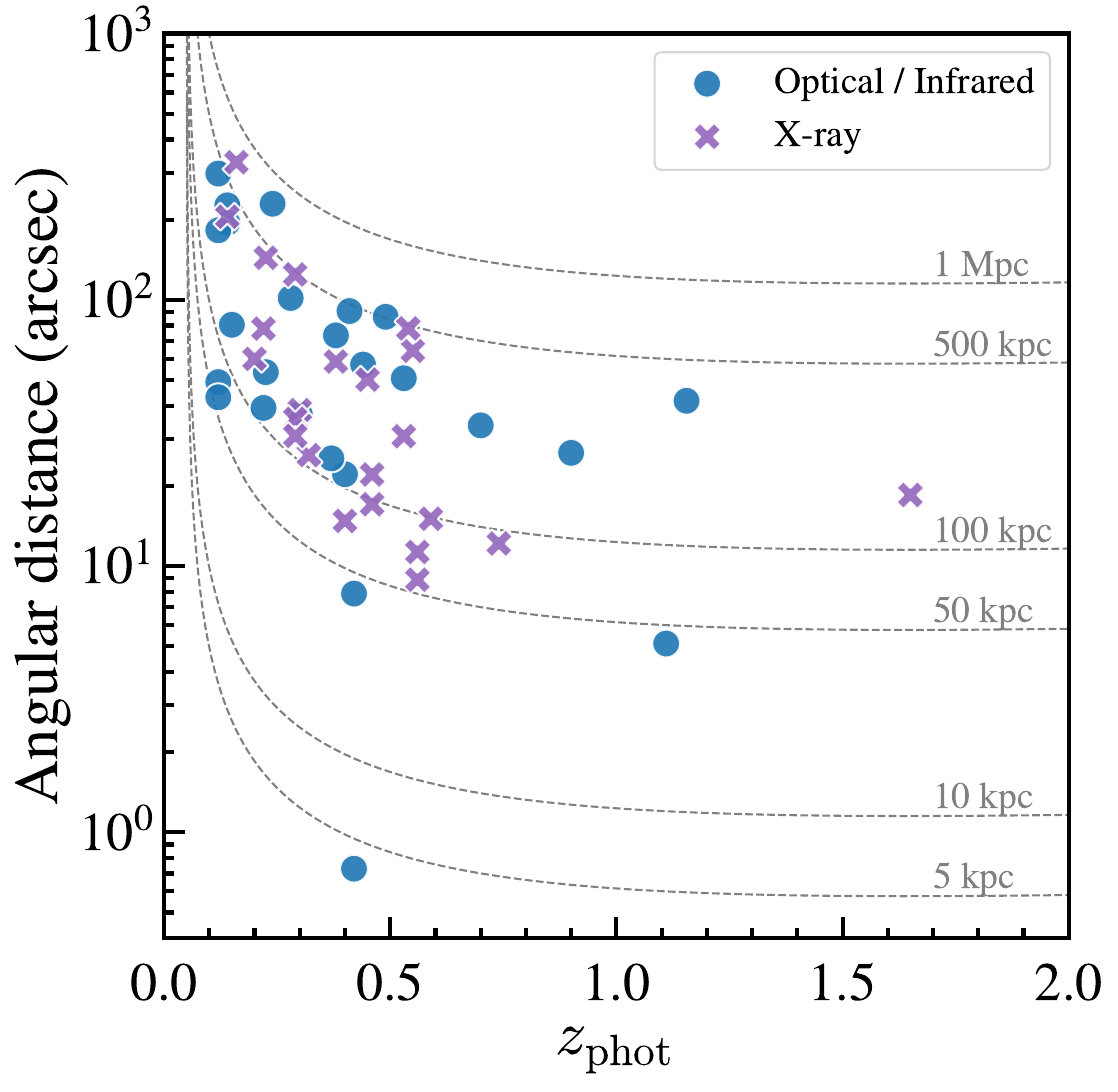}
\caption{Angular offset between IDCS cluster centers and 39 independently detected and confirmed clusters. Blue points show offsets to spectroscopically confirmed clusters from redMaPPer \citep{Rykoff2014} and other surveys. Purple crosses show offsets to X-ray detections, from XMM-Newton \citep{adami_xxl_2018} and Chandra \citep{vajgel_x-ray-selected_2014}. The dashed gray lines indicate projected physical offsets in kpc at the cluster redshift. The median offset is 211~kpc, and 79\% of matches have offsets $<0.5$~Mpc.}
\label{fig:anglexz}
\end{figure}

To further refine our analysis, we restrict our matched catalog from this point forward to include only clusters that have been independently confirmed through spectroscopic or X-ray observations.
This subset includes 57 clusters: 44 have spectroscopic confirmations from ISCS \citep{iscs_paper}, redMaPPer \citep{Rykoff2014}, and other surveys, 20 have X-ray detections from Chandra \citep{vajgel_x-ray-selected_2014} and XMM-Newton \citep{adami_xxl_2018}, with 7 clusters confirmed by both methods.
Figure~\ref{fig:anglexz} shows the angular offset between IDCS cluster positions and those of confirmed external clusters.
We exclude here the 18 confirmed IDCS clusters matched to ISCS clusters, because these datasets and cluster detection codes are not independent. For the remaining 39 independently confirmed clusters, the median offset between matched cluster centers is 211~kpc, and 79\% of the cluster positions differ by less than 0.5~Mpc. The dashed gray lines in Figure~\ref{fig:anglexz} represent projected physical offsets in kpc.

\begin{figure}[ht]
\includegraphics[width=0.95\columnwidth]{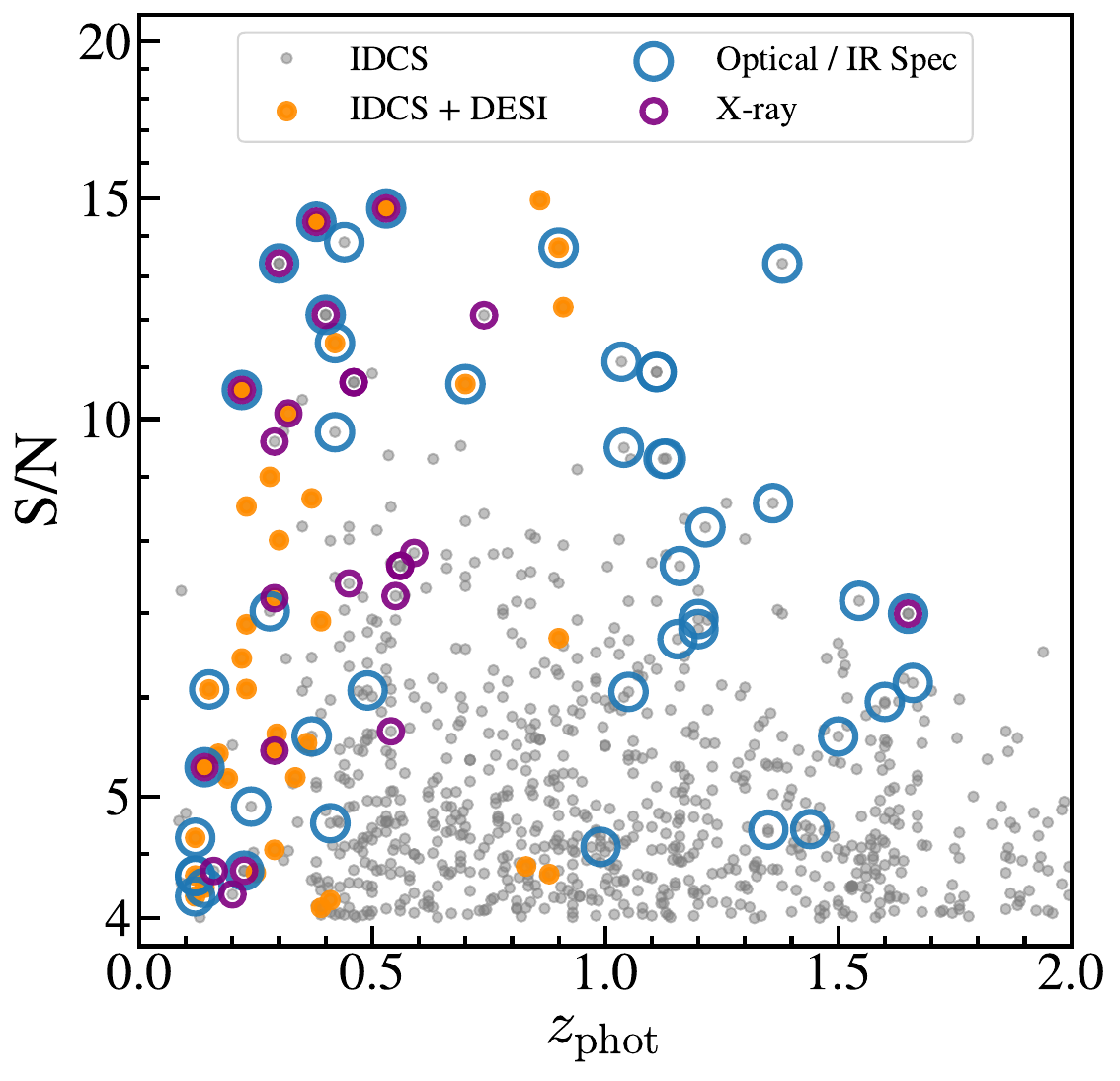}
\caption{Signal-to-noise ratio of detected clusters as a function of photometric redshift. Gray dots show all IDCS detections, with colored symbols overlaid for those with external confirmation: purple indicates clusters identified previously through X-ray; blue indicates the spectroscopically confirmed clusters; orange indicates clusters we confirmed using DESI DR1 data.}
\label{fig:zxSNR}
\end{figure}

Figure~\ref{fig:zxSNR} shows the S\slash N of detected clusters as a function of estimated redshift. At $z < 0.3$, approximately 83\% of systems with S\slash N~$>7$, 88\% with S\slash N~$>5$, and 81\% of systems with S\slash N~$>4$ are confirmed. Extending the redshift range to $z \leq 0.5$, 46 out of 136 detections (34\%) are confirmed, including 28 previously known systems and 18 newly confirmed through cross-matching with DESI galaxies. At higher redshifts ($z \geq 0.5$), only 29 out of 689 detections (4.2\%) had prior confirmation, with 5 additional systems confirmed through DESI. Notably, of the 19 systems with S\slash N~$>10$, only 2 lack external confirmation: 15 have prior X-ray or spectroscopic confirmation from the literature, and 2 additional systems were confirmed through DESI.

\begin{figure}[ht]
\includegraphics[width=0.95\columnwidth]{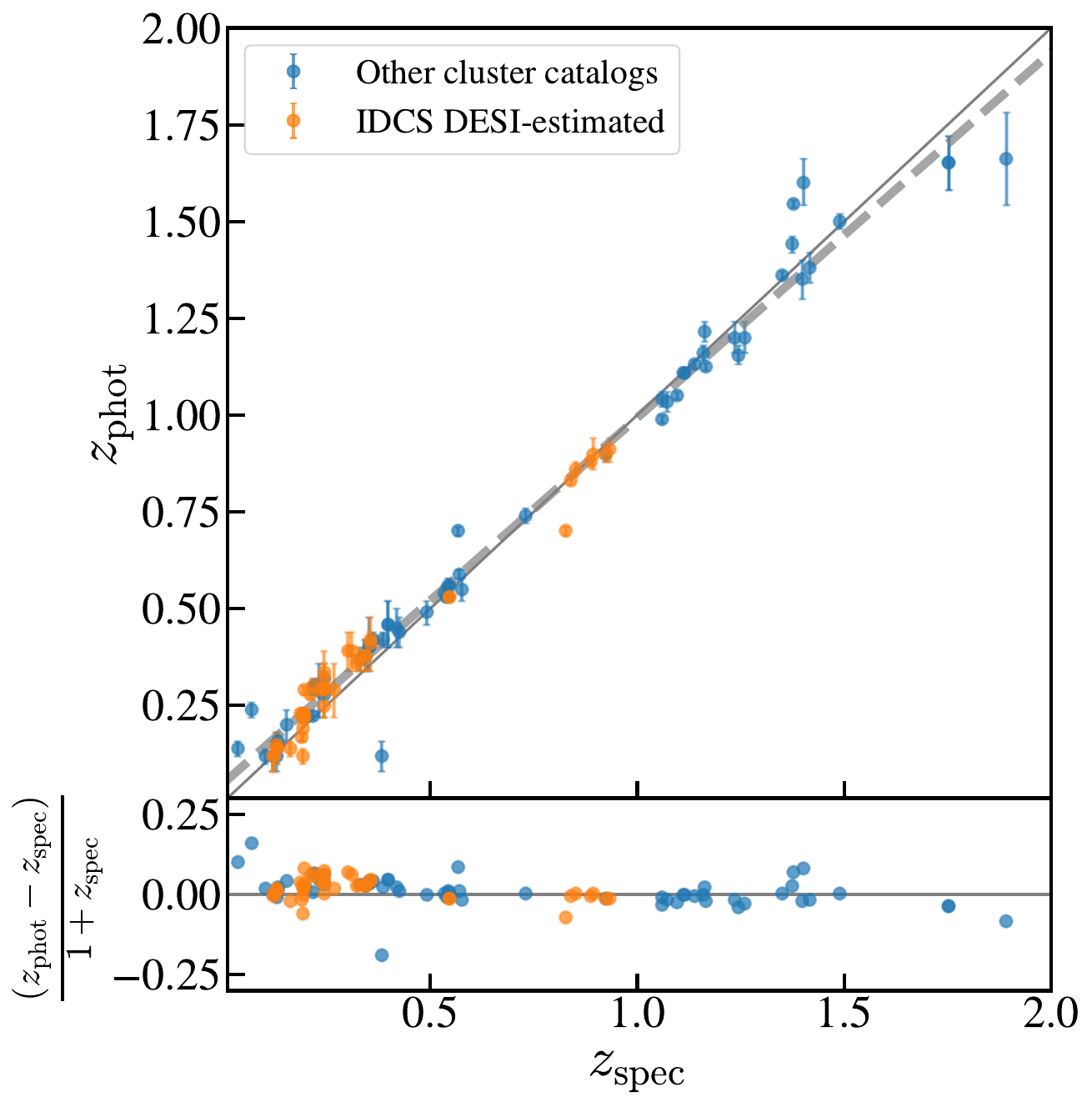}
\caption{Comparison between IDCS clusters' photometric and spectroscopic redshifts. Blue indicates spectroscopic redshifts of previously-confirmed clusters. Orange indicates spectroscopic redshifts derived using DESI DR1. The derived best-fit relation for the combined sample, which has an $\sigma_{z} / (1+z) = 0.017$, is shown as a dashed light gray line. The gray solid line is the one-to-one relation.}
\label{fig:zphotzspec}
\end{figure}

Figure \ref{fig:zphotzspec} relates our photometric redshifts to the spectroscopic redshifts of clusters that have been confirmed before and to the redshifts we infer using DESI DR1 data (see \S \ref{sec:desi}). The best fit to the points is
\begin{align}
    z_{\rm{phot,lit,init}} = (0.944 \pm 0.012) z_{\rm{spec}} + (0.048 \pm 0.009) ,
\label{eq1}
\end{align}
and we find a $\sigma_{z} = (z_{\rm{phot}}-z_{\rm{spec}})/(1+z_{\rm{spec}}) = 0.017$. The mild trend in Eq. \ref{eq1}, in which $z_{\rm phot}$ is slightly overestimated at $z \lesssim 0.8$, is consistent with the bias seen in the galaxy-level photometric redshifts in Figure~\ref{fig:photozplot}.

We note that the externally confirmed clusters shown in Figures~\ref{fig:zxSNR} and \ref{fig:zphotzspec} are relatively sparse at $0.5 < z < 1$, even though this is the most populated redshift range in the full IDCS catalog, with 312 detections. This gap reflects the different redshift ranges probed by the available external confirmation samples rather than a lack of IDCS detections. The optical and X-ray catalogs used here are concentrated primarily at lower redshifts, while the ISCS sample provides infrared-selected confirmations mainly at $z \gtrsim 1$. DESI extends into the intermediate-redshift range but is also weighted toward lower redshifts. The limited number of confirmations at $0.5 < z < 1$ therefore reflects the heterogeneous selection functions of the external samples used for validation.

\begin{deluxetable*}{lccrrcc}
\tablecaption{Biweight scale and gapper velocity dispersions, as well as redshift estimates, for IDCS clusters containing at least 10 galaxies observed by DESI DR1. \label{tab:SDWFS_and_DESI}}
\tablewidth{700pt}
\tabletypesize{\scriptsize}
\tablehead{\\ SDWFS ID & SDWFS $z$ & DESI $z$ & $\sigma_{BW}$ & $\sigma_{GAP}$ & $N_{\mathrm{gals}}$ & DESI Mass \\ 
\colhead{} & \colhead{} & \colhead{} & \colhead{$\mathrm{km\,s^{-1}}$} & 
\colhead{$\mathrm{km\,s^{-1}}$} & \colhead{} & \colhead{$10^{14}$M$_\odot$}
}
\startdata
IDCS J143116+350608 & 0.22 & 0.193 & $629 \substack{+78 \\ -32}$ & $623 \substack{+75 \\ -26}$ & 77 & $2.26 \substack{+0.83 \\ -0.25}$ \\
IDCS J142848+323748 & 0.15 & 0.129 & $447 \substack{+70 \\ -39}$ & $448 \substack{+60 \\ -41}$ & 52 & $0.95 \substack{+0.39 \\ -0.22}$ \\
IDCS J142926+340215 & 0.14 & 0.128 & $398 \substack{+56 \\ -30}$ & $393 \substack{+60 \\ -23}$ & 46 & $0.66 \substack{+0.32 \\ -0.10}$ \\
IDCS J143706+333318 & 0.32 & 0.243 & $428 \substack{+54 \\ -58}$ & $429 \substack{+59 \\ -31}$ & 42 & $0.79 \substack{+0.34 \\ -0.15}$ \\
IDCS J143233+354508 & 0.23 & 0.185 & $453 \substack{+92 \\ -33}$ & $451 \substack{+86 \\ -29}$ & 37 & $0.94 \substack{+0.58 \\ -0.16}$ \\
IDCS J142902+353713 & 0.29 & 0.235 & $525 \substack{+85 \\ -94}$ & $527 \substack{+65 \\ -75}$ & 31 & $1.40 \substack{+0.53 \\ -0.48}$ \\
IDCS J142803+345432 & 0.17 & 0.187 & $861 \substack{+115 \\ -63}$ & $763 \substack{+104 \\ -61}$ & 29 & $3.96 \substack{+1.67 \\ -0.81}$ \\
IDCS J143333+334753 & 0.12 & 0.123 & $548 \substack{+236 \\ -82}$ & $588 \substack{+139 \\ -53}$ & 28 & $2.00 \substack{+1.58 \\ -0.46}$ \\
IDCS J142453+351401 & 0.19 & 0.190 & $893 \substack{+61 \\ -74}$ & $888 \substack{+81 \\ -43}$ & 27 & $6.00 \substack{+1.63 \\ -0.76}$ \\
IDCS J142630+350855 & 0.12 & 0.190 & $724 \substack{+47 \\ -67}$ & $712 \substack{+62 \\ -56}$ & 27 & $3.27 \substack{+0.84 \\ -0.66}$ \\
IDCS J143104+323207 & 0.29 & 0.266 & $933 \substack{+149 \\ -141}$ & $884 \substack{+120 \\ -67}$ & 25 & $5.69 \substack{+2.38 \\ -1.11}$ \\
IDCS J143206+342132 & 0.22 & 0.190 & $247 \substack{+76 \\ -64}$ & $258 \substack{+62 \\ -42}$ & 23 & $0.20 \substack{+0.16 \\ -0.08}$ \\
IDCS J143300+342042 & 0.23 & 0.192 & $372 \substack{+102 \\ -47}$ & $377 \substack{+89 \\ -48}$ & 21 & $0.57 \substack{+0.45 \\ -0.18}$ \\
IDCS J142632+344319 & 0.28 & 0.211 & $436 \substack{+74 \\ -18}$ & $430 \substack{+71 \\ -18}$ & 21 & $0.81 \substack{+0.42 \\ -0.09}$ \\
IDCS J143620+334649 & 0.38 & 0.341 & $618 \substack{+243 \\ -124}$ & $648 \substack{+217 \\ -65}$ & 21 & $2.32 \substack{+2.81 \\ -0.59}$ \\
IDCS J142456+322525 & 0.23 & 0.193 & $713 \substack{+146 \\ -202}$ & $699 \substack{+97 \\ -117}$ & 20 & $3.10 \substack{+1.33 \\ -1.23}$ \\
IDCS J143539+330818 & 0.12 & 0.120 & $186 \substack{+48 \\ -22}$ & $186 \substack{+54 \\ -15}$ & 20 & $0.08 \substack{+0.09 \\ -0.02}$ \\
IDCS J143607+344715 & 0.91 & 0.931 & $819 \substack{+116 \\ -77}$ & $785 \substack{+99 \\ -87}$ & 19 & $2.78 \substack{+1.07 \\ -0.77}$ \\
IDCS J142851+350836 & 0.29 & 0.194 & $392 \substack{+77 \\ -47}$ & $387 \substack{+92 \\ -38}$ & 19 & $0.61 \substack{+0.49 \\ -0.15}$ \\
IDCS J142855+325448 & 0.30 & 0.217 & $323 \substack{+98 \\ -65}$ & $326 \substack{+76 \\ -48}$ & 18 & $0.38 \substack{+0.29 \\ -0.13}$ \\
IDCS J142646+333509 & 0.36 & 0.322 & $481 \substack{+101 \\ -91}$ & $485 \substack{+95 \\ -72}$ & 17 & $1.06 \substack{+0.67 \\ -0.38}$ \\
IDCS J142710+333801 & 0.86 & 0.850 & $295 \substack{+72 \\ -74}$ & $298 \substack{+61 \\ -67}$ & 16 & $0.20 \substack{+0.14 \\ -0.10}$ \\
IDCS J143105+353708 & 0.14 & 0.162 & $367 \substack{+65 \\ -75}$ & $355 \substack{+37 \\ -58}$ & 16 & $0.49 \substack{+0.15 \\ -0.19}$ \\
IDCS J143250+325531 & 0.29 & 0.242 & $393 \substack{+66 \\ -51}$ & $393 \substack{+69 \\ -31}$ & 16 & $0.62 \substack{+0.35 \\ -0.13}$ \\
IDCS J143834+335150 & 0.30 & 0.242 & $299 \substack{+63 \\ -56}$ & $306 \substack{+55 \\ -43}$ & 15 & $0.31 \substack{+0.18 \\ -0.11}$ \\
IDCS J142935+340740 & 0.37 & 0.327 & $608 \substack{+170 \\ -77}$ & $588 \substack{+92 \\ -76}$ & 14 & $1.79 \substack{+0.88 \\ -0.57}$ \\
IDCS J143415+331526 & 0.34 & 0.244 & $376 \substack{+129 \\ -81}$ & $390 \substack{+99 \\ -64}$ & 14 & $0.61 \substack{+0.52 \\ -0.24}$ \\
IDCS J142538+332234 & 0.25 & 0.243 & $198 \substack{+53 \\ -23}$ & $201 \substack{+38 \\ -9}$ & 13 & $0.10 \substack{+0.06 \\ -0.01}$ \\
IDCS J142441+343610 & 0.39 & 0.300 & $949 \substack{+648 \\ -764}$ & $1244 \substack{+166 \\ -181}$ & 13 & $14.27 \substack{+5.86 \\ -5.00}$ \\
IDCS J143411+345626 & 0.70 & 0.825 & $227 \substack{+77 \\ -33}$ & $232 \substack{+72 \\ -22}$ & 12 & $0.10 \substack{+0.11 \\ -0.02}$ \\
IDCS J143457+345311 & 0.90 & 0.923 & $945 \substack{+300 \\ -80}$ & $953 \substack{+246 \\ -29}$ & 12 & $4.75 \substack{+4.18 \\ -0.39}$ \\
IDCS J142452+334401 & 0.42 & 0.354 & $1118 \substack{+191 \\ -88}$ & $1114 \substack{+251 \\ -60}$ & 12 & $10.22 \substack{+7.64 \\ -1.44}$ \\
IDCS J142936+343852 & 0.39 & 0.308 & $181 \substack{+34 \\ -9}$ & $170 \substack{+34 \\ -20}$ & 11 & $0.06 \substack{+0.04 \\ -0.02}$ \\
IDCS J143713+341529 & 0.53 & 0.545 & $705 \substack{+181 \\ -49}$ & $719 \substack{+159 \\ -48}$ & 11 & $2.74 \substack{+2.01 \\ -0.47}$ \\
IDCS J143203+351656 & 0.90 & 0.893 & $957 \substack{+505 \\ -87}$ & $927 \substack{+95 \\ -56}$ & 11 & $4.48 \substack{+1.38 \\ -0.71}$ \\
IDCS J143207+331544 & 0.88 & 0.885 & $815 \substack{+150 \\ -179}$ & $821 \substack{+130 \\ -136}$ & 11 & $3.23 \substack{+1.61 \\ -1.26}$ \\
IDCS J142856+334537 & 0.83 & 0.837 & $627 \substack{+123 \\ -49}$ & $643 \substack{+114 \\ -38}$ & 10 & $1.70 \substack{+0.96 \\ -0.26}$ \\
IDCS J143542+344454 & 0.41 & 0.354 & $517 \substack{+136 \\ -46}$ & $494 \substack{+74 \\ -52}$ & 10 & $1.08 \substack{+0.51 \\ -0.31}$ \\
\enddata
\end{deluxetable*}

\begin{figure}[ht]
\includegraphics[width=0.95\columnwidth]{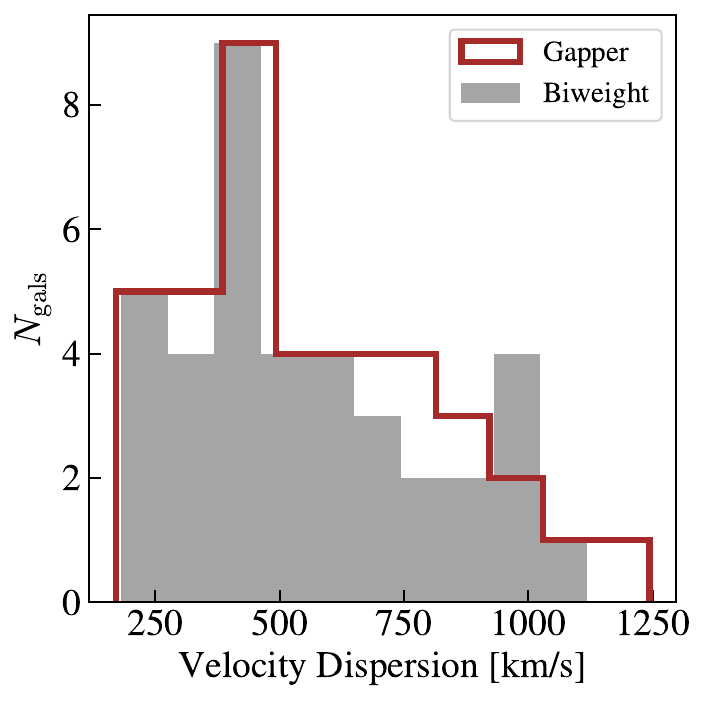}
\caption{Distribution of velocity dispersion estimates for clusters with at least 10 cluster members from DESI, using the biweight scale estimator (gray) and the gapper estimator (red).}
\label{fig:veldisp}
\end{figure}

\begin{figure}[ht]
\includegraphics[width=0.95\columnwidth]{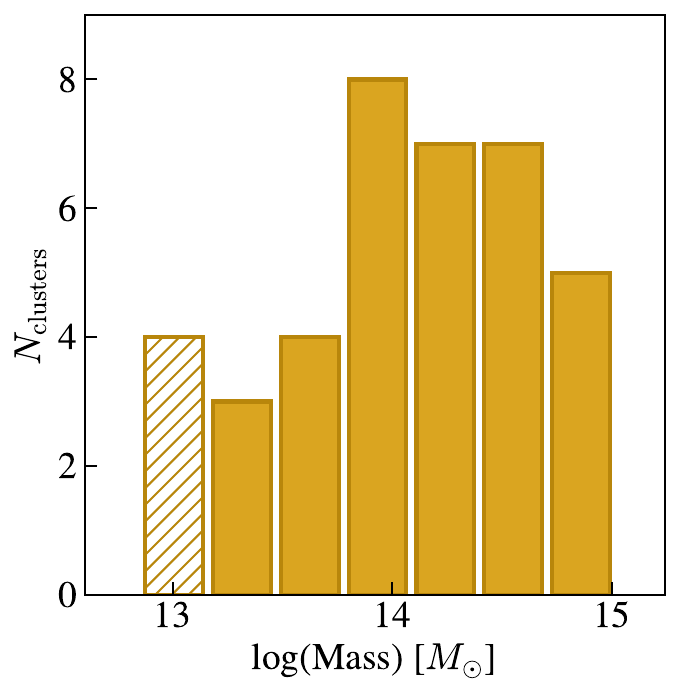}
\caption{Distribution of dispersion-based mass estimates for clusters with at least 10 member galaxies from DESI. Hatched bins indicate systems with mass below $10^{13}$, for which the \citet{munari_relation_2013} scaling relation is not well calibrated; mass estimates in this regime are therefore uncertain.}
\label{fig:masses}
\end{figure}

\subsection{Redshifts and velocity dispersion estimates with DESI} \label{sec:desi}

In addition to comparing with other catalogs, we enhance our spectroscopic reference set using DESI DR1 data. This dataset allows us to derive redshifts and velocity dispersions for a subset of clusters. We follow the guidelines in \cite{desiDR1} to clean the DESI catalog: excluding stars, and selecting galaxies with reliable redshift measurements, removing those with low confidence fits, ambiguous classifications, or poor spectral quality. For objects with multiple observations, we retain only the highest S\slash N. We also restrict the sample to galaxies within the redshift range $0<z<2$, and exclude entries with problematic photometry.

For cluster association, every DESI galaxy within 1.5 Mpc and $\Delta z = 0.1 (1+z)$ of an IDCS galaxy cluster is initially assigned to that cluster. We then retain only clusters that have at least 10 galaxies within $3\sigma$ of the mean cluster velocity. To avoid spurious associations, we also only keep clusters if the peak of the DESI spectroscopic redshift distribution, constructed from galaxies within $1.5$ Mpc of the cluster center, lies within $\Delta z=0.09 (1+z)$ of the cluster photometric redshift and is at least twice as prominent as any secondary peak in the same distribution. The peak comparison is performed on a smoothed version of the redshift histogram, using a smoothing length of $1000$ $\mathrm{km\,s^{-1}}$.

\begin{figure}[ht]
\includegraphics[width=0.95\columnwidth]{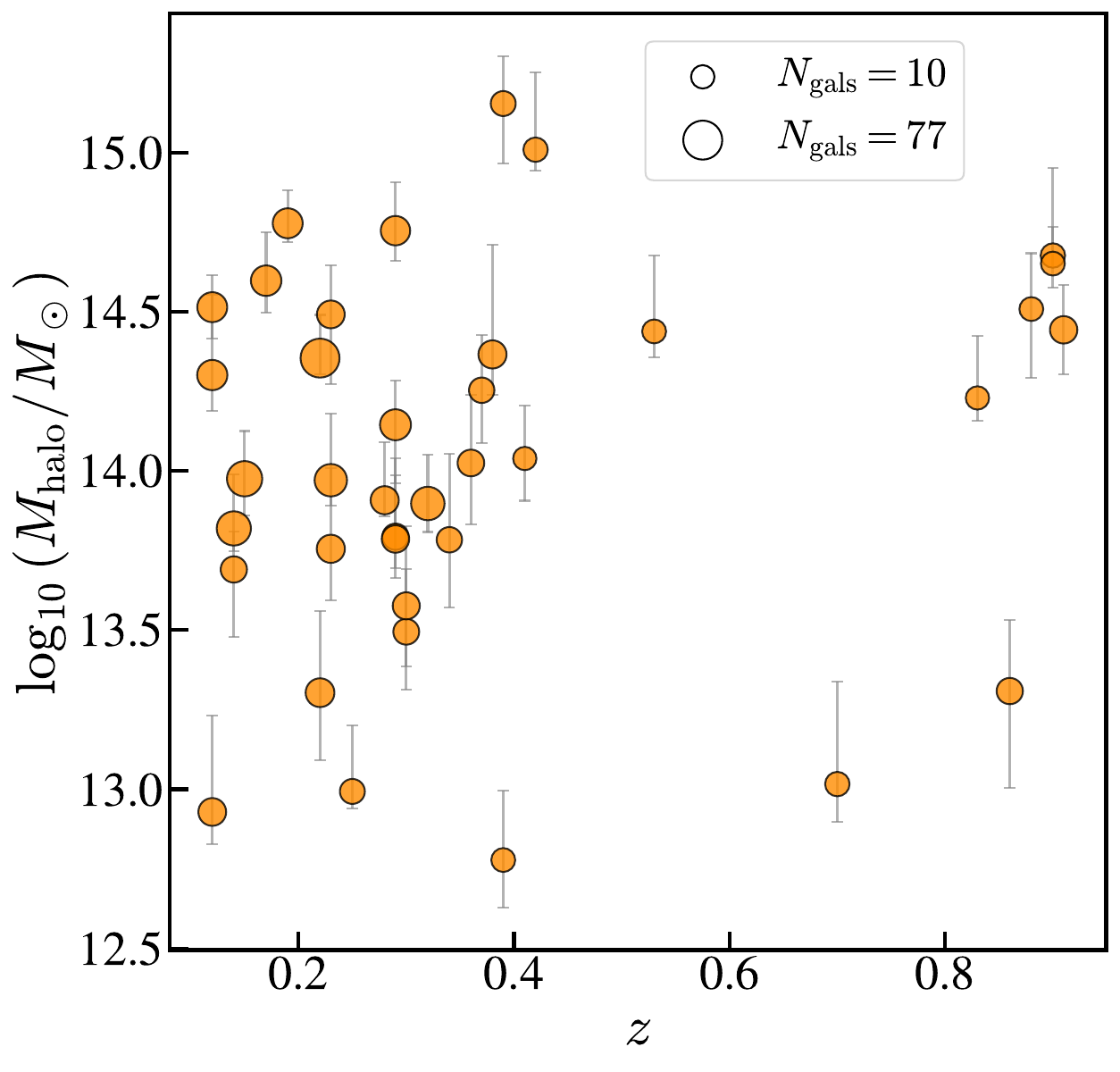}
\caption{Estimated cluster masses as a function of redshift for the 38 systems with at least 10 DESI spectroscopic members. Circle size indicates the number of galaxies used in the velocity dispersion estimate, with reference symbols in the legend marking the minimum ($N_{\rm{gals}}=10$) and maximum ($N_{\rm{gals}}=77$).}
\label{fig:mass_z_Ngal}
\end{figure}

We use the code {\tt pyrostat} \citep{RuelThesis}, adapted to Python from its original FORTRAN implementation by \citep{beers_measures_1990}, to compute velocity dispersions for IDCS clusters using the gapper ($\sigma_{\mathrm{GAP}}$) and the biweight scale variance ($\sigma_{\mathrm{BWT}}$) estimators. {\tt pyrostat} identifies the members of the cluster and estimates the cluster redshifts and velocity dispersions through a robust multi-step procedure designed to mitigate the impact of outliers and substructures.

The algorithm begins by computing the biweight location of the galaxy redshifts in each candidate cluster, which serves as an initial estimate of the cluster center in velocity space. A broad first-pass cut removes galaxies with rest-frame velocity offsets greater than $6000$ $\mathrm{km\,s^{-1}}$ from this center. This step eliminates extreme outliers and misidentified redshifts without assuming Gaussianity.

The sample is then further refined using an iterative $3\sigma$ clipping procedure based on the estimated velocity dispersion. To ensure that the final sample is not affected by lingering interlopers or local substructure, a final hard cut removes galaxies with offsets greater than $3000$ km\,s$^{-1}$ from the updated biweight location. This multilayered approach balances the need for robustness with the need to retain a sufficient number of members for statistical reliability.

The velocity dispersion is estimated from the cleaned sample using both the gapper and biweight scale estimators, as shown in Figure~\ref{fig:veldisp}. These estimators are part of a broader family of robust statistics introduced and evaluated by \citet{beers_measures_1990} for their performance under non-Gaussian conditions. Following the definitions in \citet{tukey1958} and \citet{ferragamo2020}, and interpreting the scale estimator as a measure of velocity dispersion, we define the biweight scale as:
\begin{equation}
\sigma_{\mathrm{BWT}} = \left( \frac{{N_{\mathrm{gal}}}^2}{N_{\mathrm{gal}} - 1} \right)^{1/2} 
\left[
\frac{\sum_{|u_i|<1} (x_i - M)^2 (1 - u_i^2)^4}{\left| \sum_{|u_i|<1} (1 - u_i^2)(1 - 5u_i^2) \right|}
\right]^{1/2},
\end{equation}
where 
\begin{equation}
u_i = \frac{x_i - M}{a \times \mathrm{median}(|x_i - M|)}.
\end{equation}
We adopt \( a = 9.0 \) following \citealt{ferragamo2020}, which defines the cutoff for down-weighting outliers based on their distance from the central value. The choice balances robustness to outliers with statistical efficiency, including galaxies within approximately \(\pm6\) standard deviations under a normal distribution. Additionally, \( M \) is the sample median, and \( x_i \) corresponds to the rest-frame velocity offset of the \( i \)-th galaxy. These velocity offsets are computed as:
\begin{equation}
x_i = c \, \frac{z_i - z_{\mathrm{cluster}}}{1 + z_{\mathrm{cluster}}},
\end{equation}
where \( z_i \) is the spectroscopic redshift of the \( i \)-th galaxy, \( z_{\mathrm{cluster}} \) is the estimated cluster redshift, and \( c \) is the speed of light.

The gapper estimator \citep{wainer1976}, interpreted as a measure of velocity dispersion here, is designed for small samples and based on ordered velocity gaps. The dispersion is given by:
\begin{equation}
\sigma_{\mathrm{GAP}} = \frac{\sqrt{\pi}}{N_{\mathrm{gal}}(N_{\mathrm{gal}} - 1)} \sum_{i=1}^{N_{\mathrm{gal}} - 1} w_i g_i,
\end{equation}
where the velocity offsets \( x_i \) are first sorted in ascending order, the gaps are defined as \( g_i = x_{i+1} - x_i \), and the weights are \( w_i = i(N_{\mathrm{gal}} - i) \).

These estimators outperform the standard deviation when the underlying distribution deviates from Gaussianity or contains outliers. The biweight scale is more efficient in larger, roughly symmetric samples, while the gapper is more stable for small samples of galaxies and insensitive to substructure. For both estimators, the systems span a broad range of velocity dispersions, with median values of 499 $\mathrm{km\,s^{-1}}$ for the biweight scale estimator and 490 $\mathrm{km\,s^{-1}}$ for the gapper.

The entire cleaning and estimation procedure is repeated until convergence or a maximum of 10 iterations. This yields reliable redshift and dispersion estimates for 38 out of our 825 cluster candidates that retain at least 10 galaxy members after cleaning, and these are listed in Table \ref{tab:SDWFS_and_DESI}. The final cluster redshifts ($z_{\mathrm{DESI}}$) are reported as the biweight location of the velocities of all galaxies selected as members. The statistical uncertainty on the cluster redshift is estimated by bootstrap resampling of the member velocities.

Using the velocity dispersions, we estimate the mass of this subset of clusters following \citet{munari_relation_2013}:
\begin{align}
    \frac{  M_{200}}{ \rm{M}_{\odot}} = \frac{10^{15}}{h(z)} \left( \frac{\sigma_{\rm{GAP}}}{\textit{A}} \right)^{1/\alpha} ,
\end{align}
where $M_{200}$ is the mass inside a sphere drawn from the halo center enclosing a mean density of $200 \times \rho_{c}(z)$, with $\rho_{c}(z)$ being the critical cosmic density at redshift $z$, and \( h(z) = H(z)/100\,\mathrm{km\,s^{-1}\,Mpc^{-1}} \) is the dimensionless Hubble parameter at redshift \( z \). $A$ and $\alpha$ are empirical constants calibrated from simulations to relate halo mass and velocity dispersion. We adopt $A = 1177$ km\,s$^{-1}$ and $\alpha = 0.364$, the values calibrated for galaxies in simulations with AGN feedback \citep{munari_relation_2013}.

We estimate cluster masses using velocity dispersions derived from the gapper estimator, motivated by its robust performance in the low-$N_{\mathrm{gal}}$ regime. \cite{ferragamo2020} demonstrated that the gapper estimator remains nearly unbiased and exhibits low variance down to 10 spectroscopic members, making it well suited for our sample. They also find that the gapper estimator provides stable results across a wide range of galaxy numbers and is less sensitive to interlopers and substructures than either the standard deviation or the biweight scale estimator, particularly in small samples. Since our mass estimates depend on approximately the cube of the measured velocity dispersion, minimizing statistical bias in $\sigma_v$ is crucial.

We include our mass estimates in Table~\ref{tab:SDWFS_and_DESI}, and their distribution in Figure \ref{fig:masses}. The median mass of the 38 groups and clusters characterized using DESI galaxies is $1.1 \times 10^{14}\,\mathrm{M}_{\odot}$. We note that the scaling relation from \citet{munari_relation_2013} is calibrated for systems with $M \gtrsim 10^{13}\,\mathrm{M}_{\odot}$. As such, mass estimates for the lowest-dispersion systems (represented by hatched lines in Figure \ref{fig:masses}) in our sample are uncertain and should be interpreted with caution.

Figure~\ref{fig:mass_z_Ngal} shows the estimated cluster masses as a function of redshift for the 38 systems with DESI-confirmed members, color-coded by the number of spectroscopic galaxies used. Most of the detected systems lie at lower redshifts, where DESI is more complete, but a few higher-mass clusters are also recovered at $z > 0.5$. As expected, the number of spectroscopic members tends to decrease with redshift, reflecting the increasing difficulty of confirming cluster membership as galaxies become fainter and sparser in the spectroscopic sample.

Figure~\ref{fig:imgs1} illustrates examples of our six highest S\slash N cluster detections. These visualizations help contextualize the results discussed above by showing both the SDWFS imaging and DESI spectroscopic coverage for selected systems. On the left, we display images from the DESI Legacy Imaging Surveys Data Release 9 (DR9) \citep{dey_legacy_2019}, and on the right, the corresponding Spitzer SDWFS images \citep{ashby2009}.

\section{Conclusions} \label{sec:conc}

We present the IRAC Distant Cluster Survey (IDCS), which provides a catalog of galaxy cluster and group candidates identified using the \texttt{PZWav} detection code applied to deep, multiwavelength imaging in the 9.3 deg$^2$ Bo\"otes field. Photometric redshifts are derived from twelve-band optical and infrared photometry. This work provides a cluster catalog whose selection function we characterize directly through simulations (\S\ref{sec:selecfunc}), offering a resource to calibrate and validate detection techniques for wide-area cluster surveys.

The final catalog consists of 825 cluster candidates with (S\slash N) $\geq 4$, of which 371 meet a more conservative threshold of S\slash N $\geq 5$. The survey probes clusters over the redshift range $0.08 < z < 2.0$, with a median redshift of 0.93 for the full sample, and includes 147 cluster candidates at $z_{\rm{phot}} \geq 1.5$ with S\slash N$\geq4$. Cluster photometric redshifts, estimated within the \texttt{PZWav} framework, achieve an RMS scatter $\sigma_z/(1+z)=0.017$ for cluster photometric redshifts. Of the 825 cluster candidates, 768 are new detections not previously identified in existing X-ray or optical cluster catalogs. The remaining 57 systems are matched to published clusters that are either spectroscopically confirmed or X-ray-detected clusters. A positional comparison of our \texttt{PZWav} centroids with those from the literature, excluding 18 clusters matched only to ISCS or prior IDCS detections that share the same underlying data, yields a median absolute offset of 211 kpc from 39 independently confirmed systems.

The simulation-based analysis shows that the adopted S/N $\geq 4$ threshold yields a detection purity of $93\%$ with respect to simulated halos with $M_{\rm vir} \geq 10^{13}\,M_\odot$. Fitting an error function to the completeness, the catalog reaches $50\%$ completeness at $\log(M_{200c}/M_\odot) = 13.74 \pm 0.01$ (with a width $\sigma = 0.40$) for $z < 1.5$, with an integrated completeness of $\sim\!86\%$ for $M_{200c} \geq 10^{14}\,M_\odot$. The completeness is most robustly characterized for cluster-scale halos and should be regarded as a lower limit at $z \gtrsim 1.5$, where the real galaxy catalog becomes observationally incomplete.

We further characterize our cluster sample using spectroscopic data from  DESI DR1. We identify 38 clusters with at least 10 spectroscopic member galaxies, enabling robust measurements of redshift, velocity dispersion, and dynamical mass. The median velocity dispersions for this subset are 499 $\mathrm{km\,s^{-1}}$ using the biweight scale estimator and 490 $\mathrm{km\,s^{-1}}$ using the gapper estimator. For mass estimation, we adopt the gapper-based velocity dispersions in conjunction with the scaling relation from \citet{munari_relation_2013}. This choice is motivated by the gapper estimator’s well-established robustness in the low-$N_{\mathrm{gal}}$ regime, where it remains nearly unbiased and less sensitive to interlopers or substructure compared to other estimators. Together with the 57 previously confirmed clusters from the literature (some overlapping with the DESI-measured sample), this brings the total number of spectroscopically or X-ray confirmed systems in the IDCS catalog to 80. The resulting median halo mass for the DESI-characterized clusters is $1.1 \times 10^{14}\,M_\odot$, with individual estimates ranging from $M \sim 10^{13}$ to $10^{15}\,M_\odot$, confirming that the IDCS catalog contains both group- and cluster-mass systems.

These results highlight the value of IDCS as a catalog combining wide redshift coverage with detailed characterization in a well-studied extragalactic field. The catalog includes 147 candidates at $z\geq1.5$, demonstrating the ability of deep optical and infrared photometry to identify candidate structures at redshifts where existing catalogs become increasingly sparse. The high fraction of new detections further highlights the complementarity of the IDCS selection to shallower surveys, particularly at $z>1$. As wide-field missions such as LSST, Roman, and Euclid begin delivering survey data, IDCS provides a useful benchmark for testing photometric cluster selection and redshift estimation strategies.

\section{Acknowledgements} \label{sec:acknowledge}
We acknowledge the NDWFS, SDWFS, and ISCS teams, as well as the prior IDCS studies that identified individual high-z clusters in the Boötes field. We also thank the DESI Collaboration for making its data publicly available, and Kenneth Duncan and collaborators for putting together the photometric catalog used in this work. Finally, we thank the developers of {\tt ROSTAT} (led by Timothy Beers) and {\tt pyrostat} (led by Jonathan Ruel) for making their software accessible, and Arjun Dey for helpful comments on the manuscript. Finally, we thank the anonymous referee for a constructive report that improved this work.

\begin{figure*}[htb!]
\centering
    \includegraphics[width=0.8\linewidth]{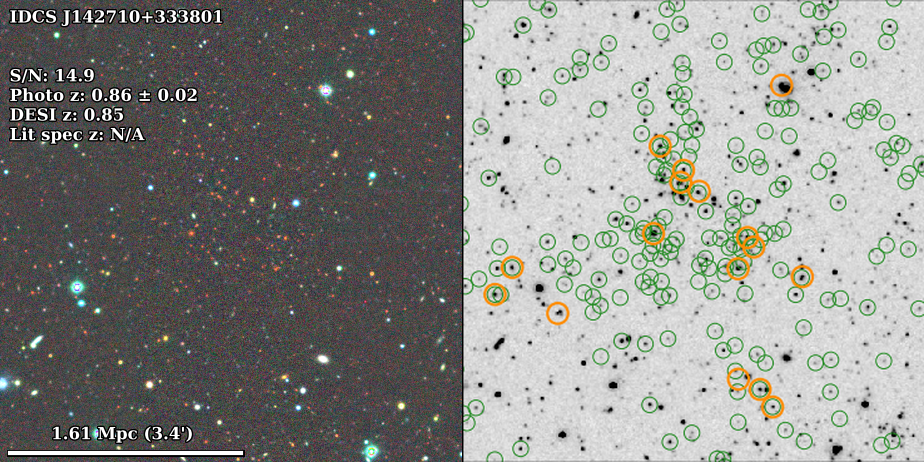}\par\medskip
    \includegraphics[width=0.8\linewidth]{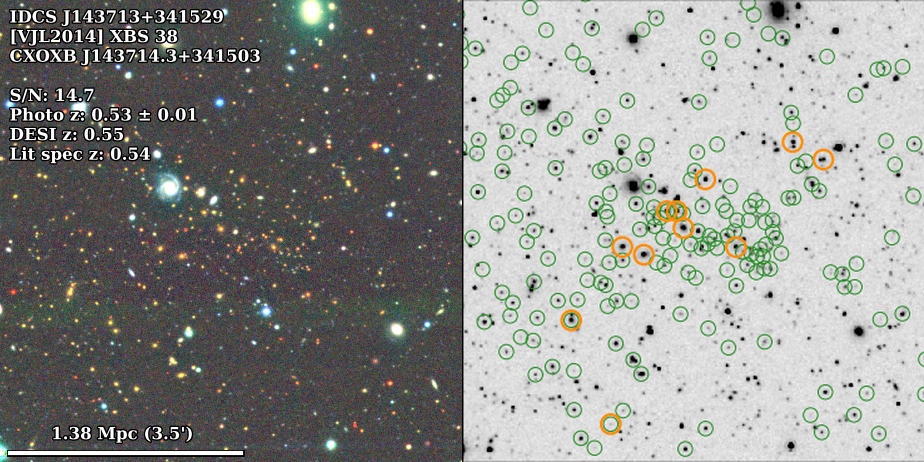}\par\medskip
    \includegraphics[width=0.8\linewidth]{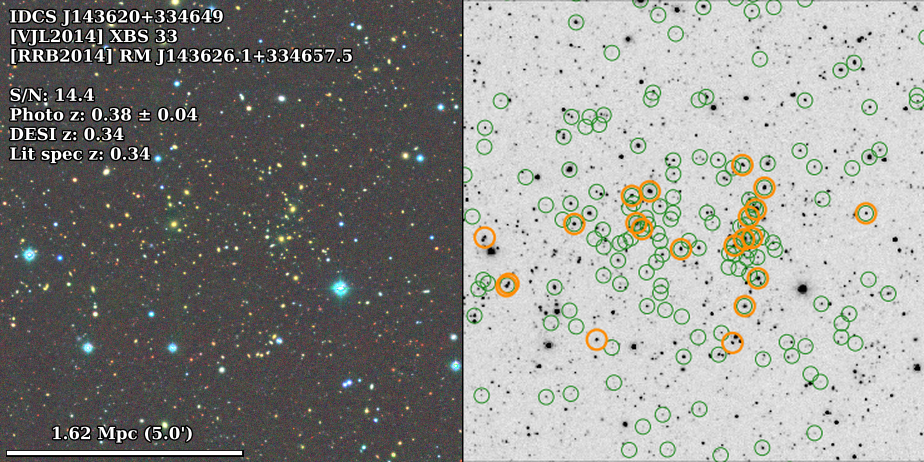}
\caption{Images from the Legacy Survey (left) and Spitzer/SDWFS (right) for the highest-S\slash N clusters in our sample that contain at least 10 DESI spectroscopic members. North is up and east is to the left. Green circles indicate galaxies from our photometric redshift catalog that lie within $0.15(1+z)$ of the cluster redshift. Orange circles mark galaxies with spectroscopic redshifts from DESI, which are used to compute velocity dispersions and masses. The upper labels on each panel list the cluster name, any known literature cross-identifications, the detection S\slash N, photometric and spectroscopic redshifts, and the DESI-estimated redshift.}
\label{fig:imgs1}
\end{figure*}

\begin{figure*}[htb!]
\addtocounter{figure}{-1}
\centering
    \includegraphics[width=0.8\linewidth]{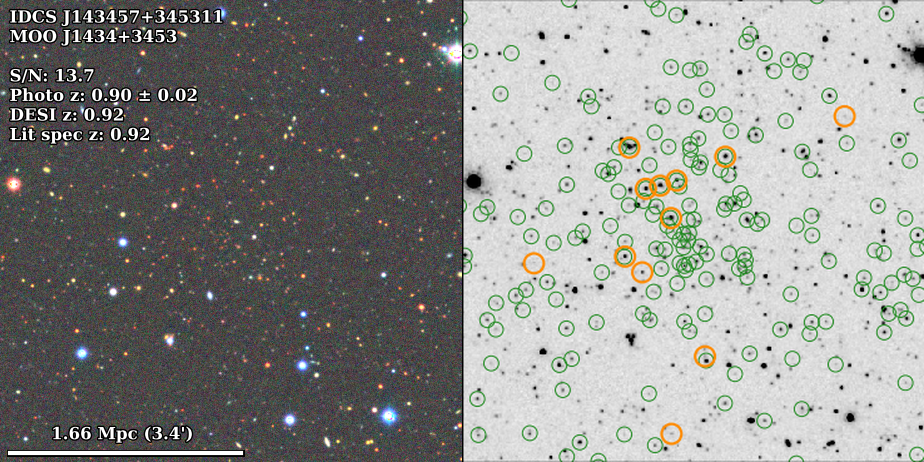}\par\medskip
    \includegraphics[width=0.8\linewidth]{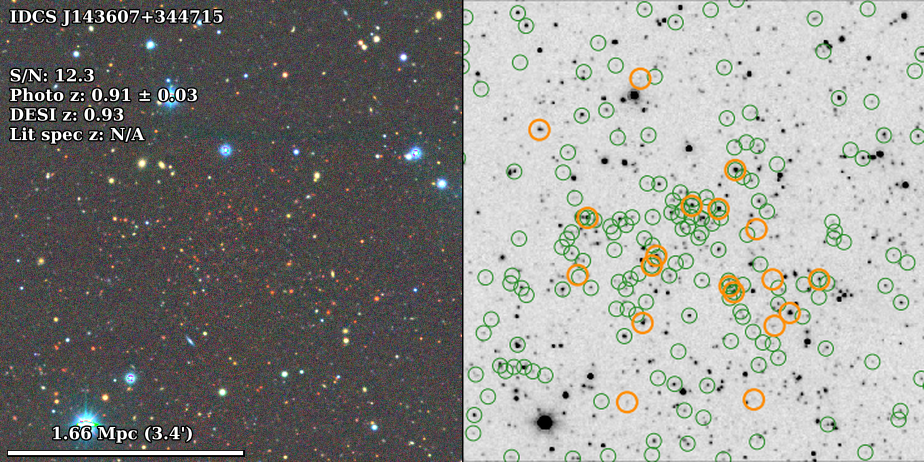}\par\medskip
    \includegraphics[width=0.8\linewidth]{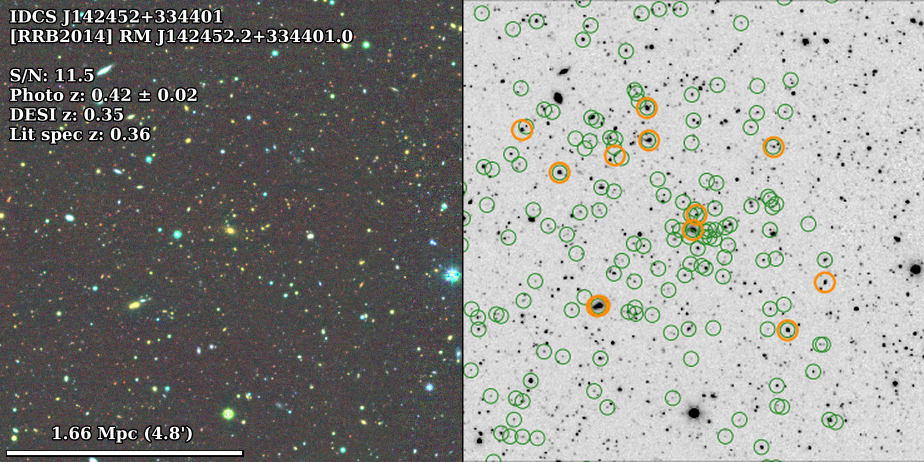}
\caption{(continued)}
\end{figure*}

\clearpage

\clearpage

\bibliography{all}{}
\bibliographystyle{aasjournalv7}

\end{document}